\documentclass[12pt, a4paper, bookmarks=false,twoside]{article}

\usepackage{times}
\usepackage{graphicx}
\usepackage{wrapfig}
\usepackage{url,color}
\usepackage{hyperref}
\usepackage{capt-of}
\usepackage{amssymb} % for lesssim

\usepackage[numbers,sort&compress]{natbib}

\definecolor{uni}{rgb}{0.490,0.604,0.667}
\definecolor{ippblue}{cmyk}{1,0.45,0.04,0.}
\definecolor{mygrey}{gray}{0.30}
\definecolor{mygreen}{rgb}{0.14,0.55, 0.14}

\newcommand{\hagis}{\textsc{Hagis}}
\newcommand{\ligka}{\textsc{Ligka}}
\newcommand{\ascot}{\textsc{Ascot}}
\newcommand{\hmt}{\textsc{Hamiltonian Mapping Technique}}
\newcommand{\asdex}{\textsc{Asdex}}
\newcommand{\iter}{\textsc{iter}}

\newcommand{\fref}{fig.\ \ref}    % abbreviation
\newcommand{\Fref}{Fig.\ \ref}    % abbreviation

\newcommand{\sref}{sec.\ \ref}    % abbreviation
\newcommand{\mr}{\mathrm}

\hypersetup{pdftitle={Nonlinear Energetic Particle Transport in the Presence of Multiple Alfv{\'e}nic Waves in ITER},
            pdfauthor={Mirjam Schneller}, % sets owner to meta data (copy right)
            pdfsubject={Paper EPS 2015},
            pdfcreator={\LaTeX2e}, colorlinks, linkcolor=mygrey, 
            urlcolor=ippblue, citecolor=mygreen}

\title{Nonlinear Energetic Particle Transport in the Presence of Multiple Alfv{\'e}nic Waves in ITER}
\author{M. Schneller$^1$, Ph. Lauber$^1$, S. Briguglio$^2$\\
$^1$ \small{Max-Planck-Institut f{\"u}r Plasmaphysik, Boltzmannstr. 2, 85748 Garching, Germany}\\
$^2$ \small{ENEA, Centro Ricerche Frascati, C.P.\ 65, 00044 Frascati, Rome, Italy}}
\date{}

\begin{document}

\maketitle

\begin{abstract}
{This work presents the results of a multi mode \iter\ study on Toroidal Alfv{\'e}n Eigenmodes, using the nonlinear hybrid \hagis-\ligka\ model. It is found that main conclusions from earlier studies of \asdex\ Upgrade discharges can be transferred to the \iter\ scenario: global, nonlinear effects are crucial for the evolution of the multi mode scenario. This work focuses on the \iter\ 15 MA baseline scenario with with a safety factor at the magnetic axis of $q_0 = 0.986$. The least damped eigenmodes of the system are identified with the gyrokinetic, non-perturbative \ligka\ solver, concerning mode structure, frequency and damping. Taking into account all weakly damped modes that can be identified linearly, nonlinear simulations with \hagis\ reveal strong multi mode behavior: while in some parameter range, quasi-linear estimates turn out to be reasonable approximations for the nonlinearly relaxed energetic particle profile, under certain conditions low-$n$ TAE branches can be excited.  As a consequence, not only grow amplitudes of all modes to (up to orders of magnitude) higher values compared to the single mode cases but also, strong redistribution is triggered in the outer radial area between $\sqrt{\hat{\rho}_{\mr{pol}}} = 0.6$ and $0.85$, far above quasi-linear estimates.}
\end{abstract}

\section{Introduction}
The excitation of global instabilities by super-thermal particles in hot plasmas and the related transport processes are of great interest for the fusion community, due to their importance for burning fusion plasmas. These energetic particles (EPs) are present in magnetic fusion devices due to external plasma heating and eventually due to fusion born $\alpha$ particles. It is necessary that the super-thermal particles are well confined while they transfer their energy to the background plasma. EPs are typically super-Alfv\'enic and can destabilise Shear-Alfv\'en, Alfv\'en-acoustic waves or other global plasma modes by resonant wave-particle interaction (inverse Landau-damping). The resulting nonlinear EP transport processes from the core to the edge and the consequential particle losses reduce the plasma heating and the fusion reaction rate. In addition, the EP losses may cause severe damages to the first wall of the device.\\
Within the last years, significant advances on both experimental and theoretical side have been made leading to a more detailed understanding of EP driven instabilities. On the theoretical side, models have advanced from fluid models for the plasma background to a fully kinetic model for all the plasma species, i.e.\ background ions and electrons, as well as EPs \cite{Lauber07}. This more accurate treatment of the background leads to changes in the linear mode properties such as frequency, damping/growth and mode structure, and can also influence the nonlinear dynamics. Since the new physics that is accessible due to a more comprehensive model can directly be validated with experimental data from present-day machines, predictions for several \iter\ scenarios can be attempted.\\
In the \iter\ 15 MA scenario \cite{Polevoi02}, a ``sea'' of small-amplitude perturbations is likely \cite{GorelenkovRev14,LauberVarenna14}. This work cannot yet provide a realistic prediction for nonlinear multi mode EP transport in this \iter\ scenario. It should rather be understood as a necessary first step towards this demanding goal, since it demonstrates a physical effect that cannot be neglected for \iter: the role of linearly stable modes within the TAE spectrum. The mechanism described in the following is investigated with the hybrid \hagis-\ligka\ model. The nonlinear \hagis\ code \cite{Pinches98} treats the EPs drift-kinetically but obtains the non-perturbative mode structures, frequencies and damping from the gyrokinetic eigenvalue solver \ligka\ \cite{Lauber07}. Recent realistic modeling of double mode scenarios with the \hagis-\ligka\ code \cite{Schneller13} not only reproduced experimentally measured EP losses in \asdex\ Upgrade \cite{Garcia10}, but also revealed the importance of linearly sub-dominant modes as well as the detailed mode structures. Hence, the crucial question for the \iter\ scenario arises, if the interaction between the ``sea'' of perturbations with the EPs will drive linearly subdominant modes unstable such that EP transport occurs in a \textit{domino effect} caused by the overlap of resonances \cite{Berk95-III,Schneller13}. In this case, even modes localized outside the region of strong EP drive could be excited nonlinearly by EP transport from more core-localized modes. As a consequence, gradient depletion and EP redistribution can exceed the quasi-linear estimates. If so, the preconditions ($q$ profile, background density profile) as well as the consequences (e.g.\ losses) must be investigated to know if and how these conditions can be avoided.

\section{The HAGIS-LIGKA Model}\label{sec:Hagisligka}
The nonlinear multi scale problem of wave-particle interaction cannot be solved in a fully consistent way so far. Although recently, linear as well as first nonlinear results for single and multi mode cases were obtained with global codes originally designed for turbulence studies \cite{Zarzoso14,Dumont13}, the large computational effort for these approaches limits the possibilities of extended, realistic simulation studies. The joint Culham-IPP code project \hagis\  \cite{Pinches98} therefore follows a hybrid approach (\fref{hagisligka}): the EP distribution is evolved in a driftkinetic model and the wave contribution enters the problem via a set of pre-calculated modes. This way, the EP nonlinearities are kept, but the MHD-nonlinearities are dropped -- only the energy transfer between waves and the particles is accounted for. This leads to a redistribution of the EP population in phase space and to the self-consistent evolution of the amplitudes of (multiple) modes and their real frequencies\footnote{in this work, if the wave evolution is not fixed, nor prescribed, it will be referred to as ``self-consistent'' mode evolution.}. The wave structures and also their damping is kept fixed during a simulation. Saturation is reached in the nonlinear stage due to the local flattening of the driving gradient in the radial EP distribution. The stochastization of the EP orbits caused by overlapping resonances with different modes influences the saturation level.\\
\begin{center}
\begin{minipage}{1\textwidth}
  \centering
  \includegraphics[width=0.65\textwidth, height=4.3cm]{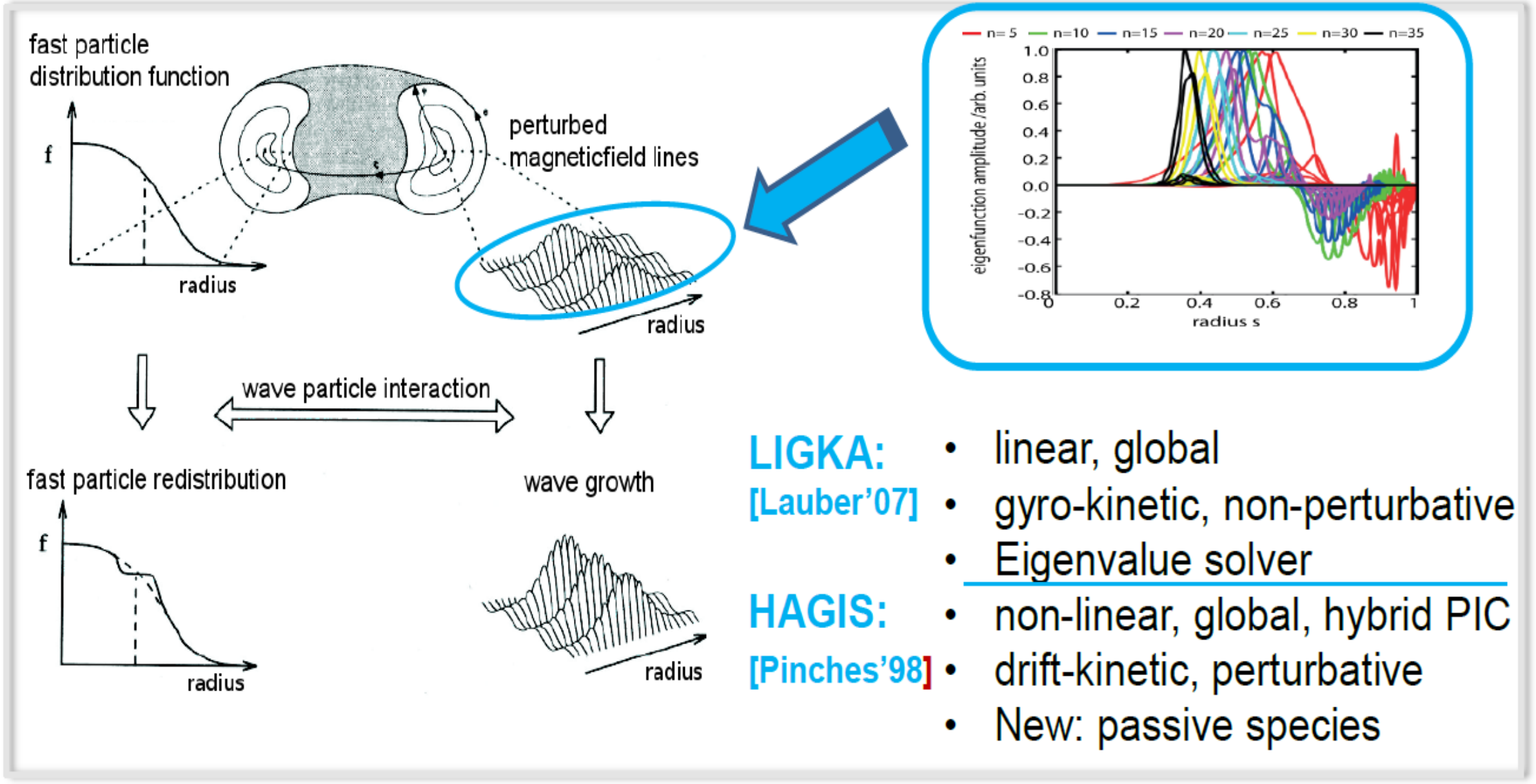}
  \captionof{figure}{\itshape \small Schematic presentation of the \hagis-\ligka\ model.}
  \label{hagisligka}
\end{minipage}
\end{center}
In ref.\ \cite{Todo10} it was found that wave-wave nonlinearities, which excite  zonal structures, can lead to steeper EP profiles for high amplitude cases. This stabilizing effect is not captured in the present model. Further, it is assumed that the absence of sources, sinks and collisions in the model is not crucial for the effect that will be shown in this work, but makes it easier to clearly demonstrate the underlying physics. The timescale of the observed nonlinear dynamics is a few milliseconds ($<$ simulation time of $\lesssim 10 ms$), which is well below plasma heating and slowing-down times ($\sim 100$ ms). Therefore, it is expected that the timescales are separated well enough to neglect the slower processes within this physics-oriented study, and additional effects of sources, sinks and collisions may alter the qualitative results only mildly.\\
Mode frequencies, damping and structures that enter \hagis\ are pre-calculated by the \ligka\ code. The ability to predict the stability of EP driven Alfv{\'e}n eigenmodes requires a detailed understanding of the dissipative mechanisms that damp these modes. To cover also the important dissipation mechanism of large scale MHD modes coupling to gyroradius scale-length kinetic Alfv{\'e}n waves, a gyrokinetic description is necessary. With the linear gyrokinetic, electromagnetic and non-perturbative code \ligka\ \cite{Lauber07} not only growth rate and damping are calculated, but also the global mode structure of the magnetic perturbation in realistic geometry due to the EPs and background kinetic effects.

\section{ITER Simulation Conditions}\label{sec:conditions}
\begin{wrapfigure}[26]{r}{0.4\textwidth}
  \includegraphics[width=0.39\textwidth]{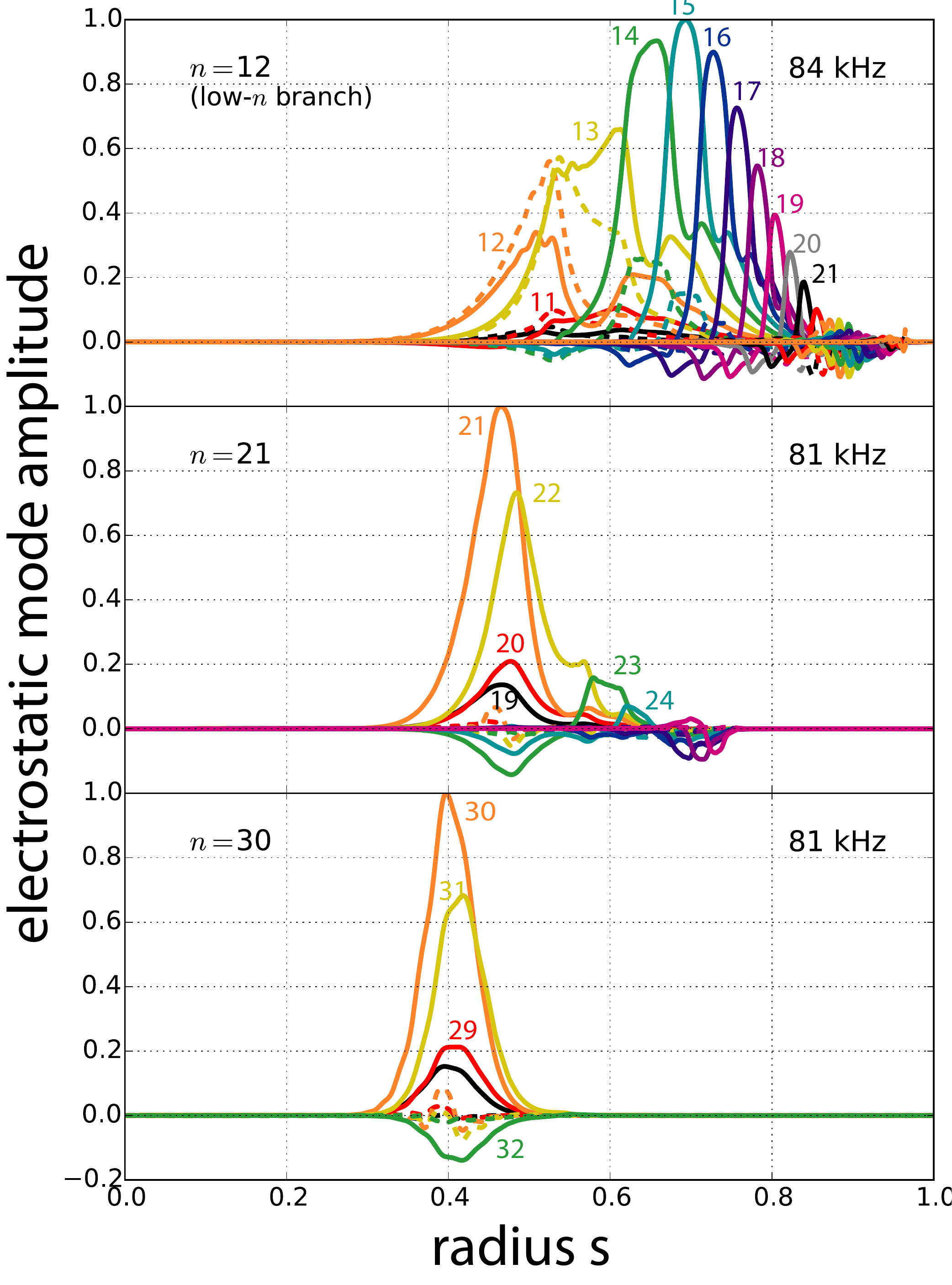}
  \caption{\itshape \small Examples of \ligka\ calculated radial structures of the electrostatic wave potential for the $n=12$ of the low-$n$ branch, $n=21$ and $n=30$ TAE in the \iter\ 15 MA scenario with $q_0=0.986$. The annotated integers denote the number of poloidal harmonic $m$.}
  \label{ITER_EF_n12+21+30}
\end{wrapfigure}

The \iter\ 15 MA baseline scenario has been analyzed in detail, presented in \cite{LauberVarenna14,Pinches15}. The findings which are relevant for this work are shortly summarized in this section: \fref{SAW-spectrum} shows the kinetic shear-Alfv\'en-wave continuum for different toroidal mode numbers $n$ and the $q$ profile as shown in \fref{qprofile} ($q_0 = 0.986$). The Toroidal Alfv{\'e}n Eigenmode (TAE) gaps are closed around\footnote{$s$ is the square root of the normalized poloidal flux.} $s \approx 0.85$. Due to the flatness of the $q$ profile and the fact that $q$ is very close to 1, the radial TAE positions $q_{\mr{TAE}} = (m+1/2)/n$ decrease monotonously with the mode number $n$ from $s \approx 0.7$ to $s\approx 0.35$ (magenta line in \fref{ITER-15MA-q099_freq}) and cluster relatively densely in the radial direction. The resonance overlap leads to additional nonlinear effects, which will be investigated in the subsequent sections. \Fref{ITER_EF_n12+21+30} shows the radial structure of the electrostatic field amplitude for three representative waves as calculated by \ligka\ and applied in the \hagis\ study of this work.\\
\begin{minipage}{0.35\textwidth}
  \includegraphics[width=1\textwidth]{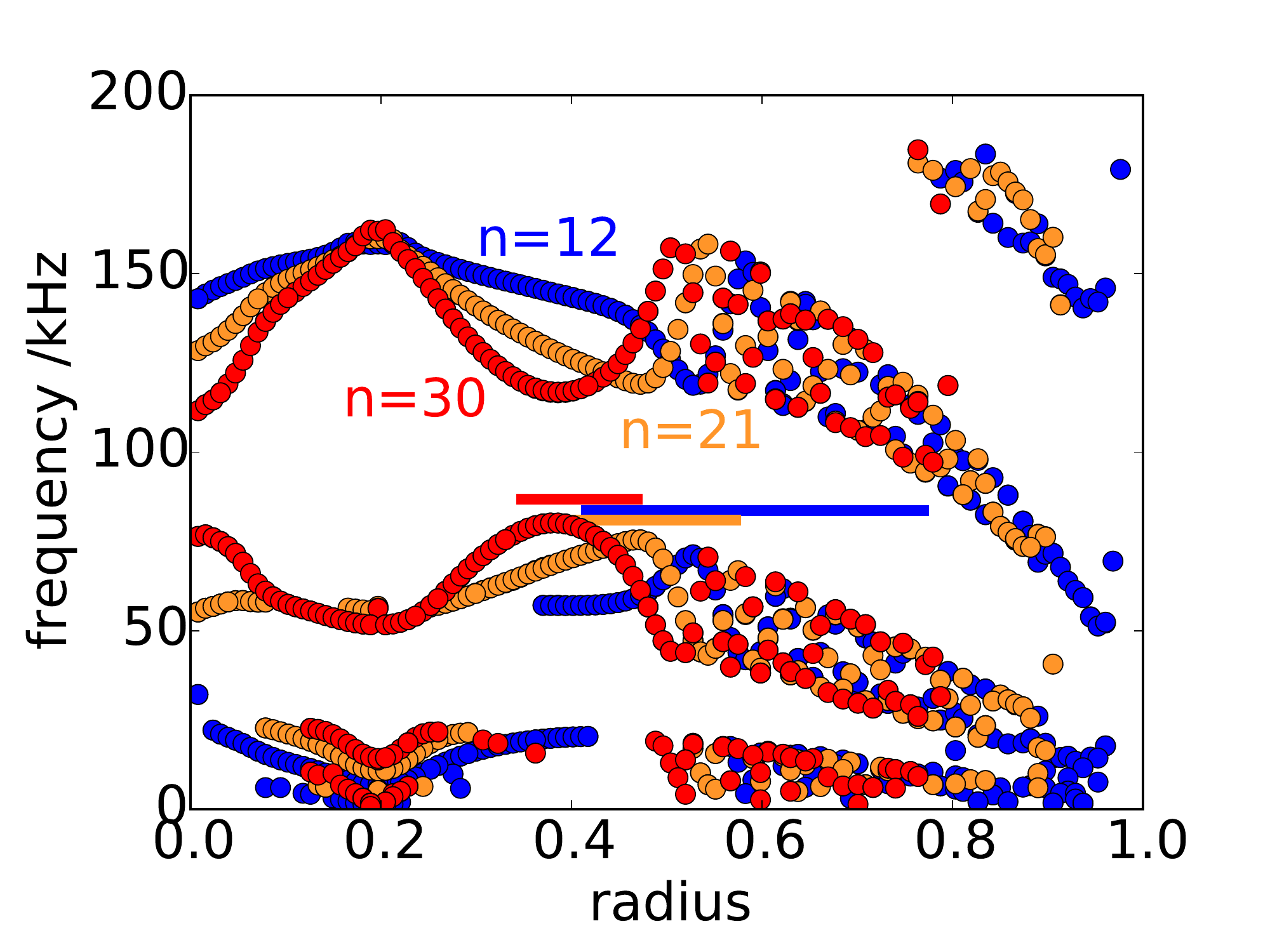}
  \captionof{figure}{\itshape \small \ligka\ calculated kinetic shear-Alfv\'en wave continuum (SAW) for the modes $n=12$ (low-$n$ branch), $21,30$ in the \iter\ 15 MA scenario with the $q$ profile as shown in \fref{qprofile} \cite{LauberVarenna14}.}
  \label{SAW-spectrum}
\end{minipage}
\hfill\begin{minipage}{0.31\textwidth}
  \includegraphics[width=1\textwidth]{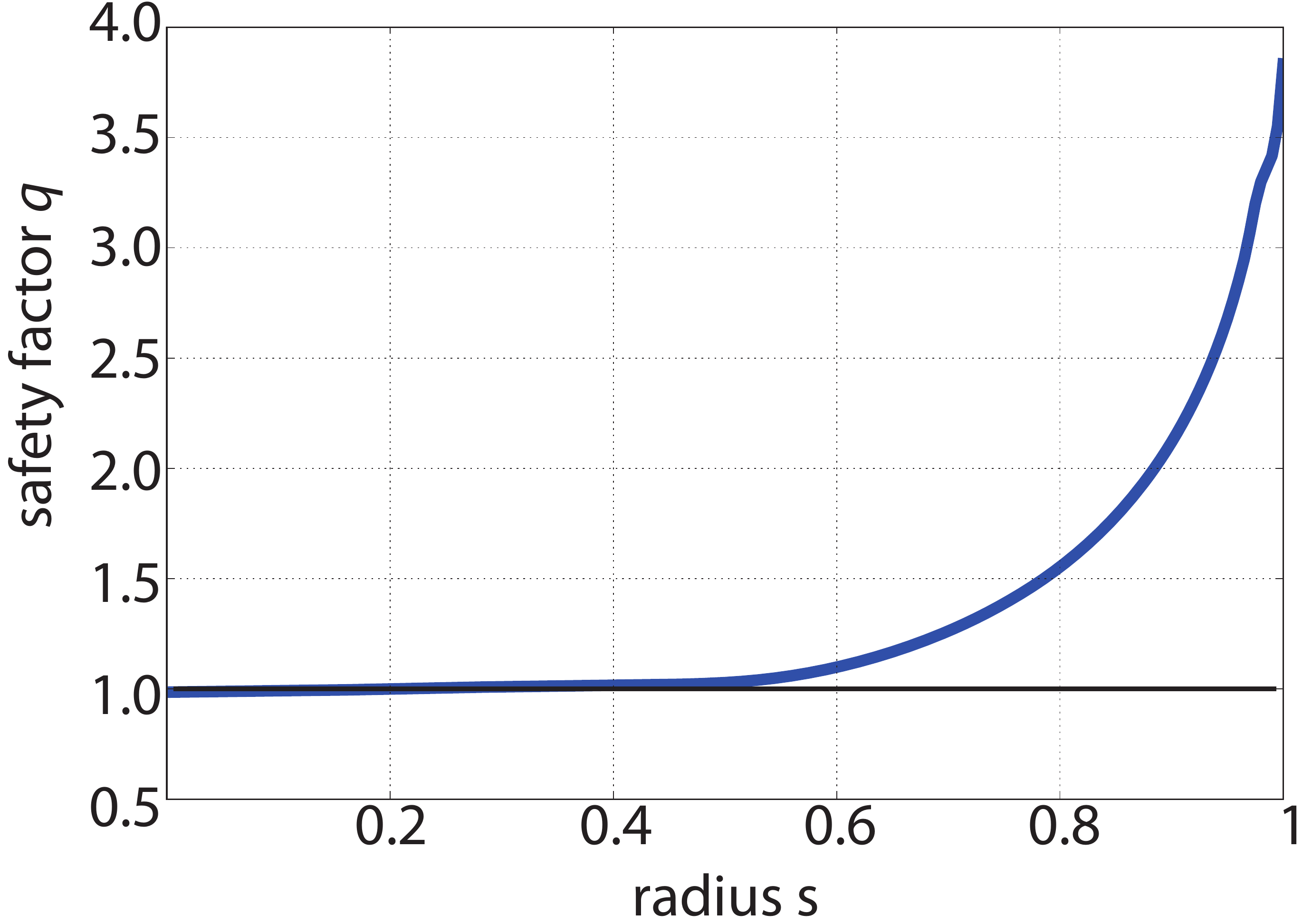}
  \captionof{figure}{\itshape \small $q$ profile used in the \iter\ simulations of this work. $q_0 = 0.986$.}
  \label{qprofile}
  \vfill
\end{minipage}
\hfill\begin{minipage}{0.3\textwidth}
  \includegraphics[width=1\textwidth]{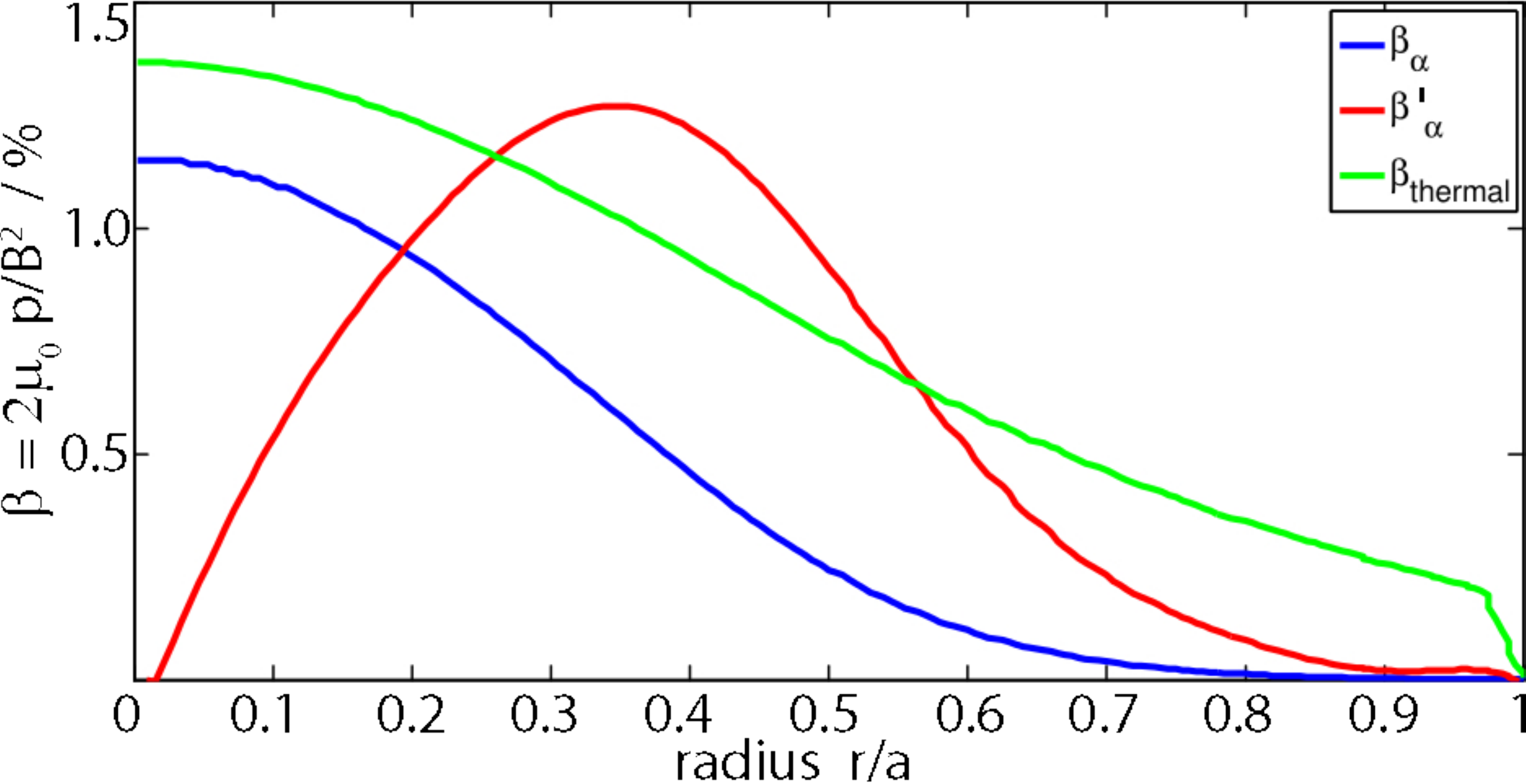}
  \captionof{figure}{\itshape \small Radial profile of EP (fusion-born $\alpha$ particles, blue) and thermal (green) $\beta$ in the presented \iter\ scenario. Red gives the radial derivative of the EP $\beta$. From \cite{Pinches15,Polevoi02}}
  \label{radialdisb}
  \vfill
\end{minipage}
The modeled EP population consists of fusion-born $\alpha$ particles, i.e.\ distributed isotropic in pitch ($\lambda$). \Fref{radialdisb} shows the EP radial distribution function, which was taken from the \iter\ database. For the energy ($E$) distribution,  a slowing-down function \cite{Gaffey76} has been assumed, with $\Delta E=491$\ keV, $E_c = 816$\ keV, $E_0 = 3.5$\ MeV:
   \begin{equation}\label{slowdown}
      f(E) = \frac{1}{E^{3/2}+E^{3/2}_{\mr{c}}}\mr{erfc}\left(\frac{E-E_0}{\Delta E}\right).
    \end{equation}
At this stage, the NBI EP population is neglected, as discussed later. Effects due to NBI ions are subject of future investigations. Further, it should be noted, that the applied distribution function is assumed to be separable in $s,E$ and $\lambda$, i.e.\ $f(s)\cdot f(E)\cdot f(\lambda)$ instead of $f(s,E,\lambda)$.
Due to the high number of modes with many poloidal harmonics, the computational effort for a nonlinear study is challenging, even with a relatively low-cost hybrid model. Together with the long time scales which have to be observed due to generally low drive in the marginally stable regime, this leads to CPU-intensive simulations. In addition, the combination of a large machine size with high toroidal mode numbers $n$ broadens the scales of resolution needed in the simulation. As a PIC code, \hagis\  resolution is determined by the number of markers. Convergence tests reveal that the necessary number of markers depends strongly on the scenario, e.g.\ on the radial distance and $n$ range of the relevant modes and can exceed 10 million for unfavorable constellations, up to 20 million for the 27-modes scenarios presented in \sref{sec:Hagisresults}.

\section{Linear Stability Analysis of the Alfv\'en Eigenmodes}\label{sec:Ligkaresults}
A detailed local and global stability analysis has been performed for the \iter\ 15 MA baseline scenario \cite{LauberVarenna14}. This scenario is not only one of the \iter\ baseline scenarios, also, it has been chosen by the EP Topical Group of the International Tokamak Physics Activity as a benchmark reference case for the study of EP behaviour in \iter. In the following, the results for the $q_0 =0.986$ case which are relevant for the nonlinear study presented in \sref{sec:Hagisresults}  are summarized. This specific $q$ profile case was chosen, since it represents a ``worst-case'' scenario in the sense that it leads to the most unfavorable constellation of gaps for weakly-damped TAE and $\alpha$ particle drive \cite{Pinches15}. However, the results which will be shown here cannot be easily compared with earlier stability analysis (such as presented in ref.\ \cite{Gorelenkov03,Gorelenkov05}, using \textsc{Nova-K} or in ref.\ \cite{ChenYang10}, using \textsc{Gem}) because the radial profiles and background magnetic equilibria are not the same. In ref.\ \cite{ChenYang10} an analytic $q$ profile is studied, which is much less box-like. This leads to smaller radial mode structures and as a consequence, the strongest mode drive (due to mode width comparable to orbit width) is found between $n=10$ and $n=20$. A recent (nonlinear) study of this \iter\ scenario is presented in ref.\ \cite{Todo14}, using the \textsc{Mega} code. There, relevant modes are reported only for $n \leq 10$, which disagrees with the findings of this work. The reasons are still to be investigated. The linear stability analysis as reported in the present section agrees well with findings of ref.\ \cite{Rodrigues15}, using a \textsc{Mishka -- Castor-K} hybrid model, as well as with qualitative analytical estimates \cite{Pinches15}.\\

\newpage
The \ligka\ eigenvalue solver finds relevant eigenmodes up to $n\approx 40$. For intermediate and low $n$, several weakly damped branches of TAE appear in the gap\footnote{The damping is named $\gamma_{\mr{d}}$, while $\gamma_{\mr{L}}$ denotes the linear mode growth without any damping. The effective growth rate is $\gamma = \gamma_{\mr{L}}-\gamma_{\mr{d}}$. Note that a convention to calculate the mode growth in ``\%'' is to normalize $\gamma$ in units 1/s by the mode frequency $\omega$ in rad/s. This definition is valid for all $\gamma$ throughout this article.}: $\gamma_{\mr{d}} \lesssim 1.0 \%$ for the low-$n$ (blue line in \fref{ITER-15MA-q099_damp}) and $\gamma_{\mr{d}} \lesssim 1.4 \%$ for the intermediate-$n$ branch (green line in \fref{ITER-15MA-q099_damp}). The main (red line in \fref{ITER-15MA-q099_damp}) branch is characterized by the two main poloidal harmonics $m$ of the mode being $m=n,n+1$, whereas for the intermediate-$n$ branch, it is $m=n+1,n+2$ and the low-$n$ branch, it is $m=n+2,n+3$. With increasing $n$, the TAE modes become more localized, whereas especially the outer, low-$n$ modes have a large number of radially extended poloidal harmonics. This effect is due to the shape of TAE gap towards the edge in the SAW. For the calculation of the mode strucutures, the $\alpha$ particle drive has been taken into account. However, the effect becomes slightly relevant only in the outer core region, modifying the coupling of the harmonics through the TAE gap. It should be noted here again, that the mode structures will be kept fixed during the investigation presented in the next section. The possibility of evolving the mode structures with time is about to be implemented currently, however, for the present investigation of marginally unstable modes, major changes in the mode structures are not expected.\\
Although the TAE position moves inwards, i.e.\ towards higher ion background temperature with increasing $n$, the damping decreases (\fref{ITER-15MA-q099_damp}). Two effects are responsible for this damping behavior: the frequency increases (\fref{ITER-15MA-q099_freq}) due to\footnote{$\omega_\mr{A}$ denotes the Alfv{\'e}n frequency  at the mag.\ axis: $\omega_\mr{A}:=v_\mr{A}/R$, where $v_\mr{A}$ is the Alfv{\'e}n velocity and R the major plasma radius (both at mag.\ axis $s=0$). In the presented scenario, $\omega_\mr{A} = 178$ kHz.} $\omega_\mr{TAE}/\omega_\mr{A} = n/(2m+1)$ (which decreases ion Landau damping) and the diamagnetic effects become more important with increasing mode numbers. Further, the modes move into the low-shear region, which decreases radiative damping. This effect can compensate for the increase of radiative damping (via $k_\perp \rho_i$) for the more localized mode structures. The least damped modes (of the higher-$n$ main branch) are found around $n=27$.\\
Adding the $\alpha$ particle population, a slight destabilization is found linearly ($\gamma < 1.5\%$) for the TAEs between $n=20$ and $35$, as can be combined from \fref{ITER-15MA-q099_damp} and \fref{ITER-gamma_n}. This is in agreement with linear \hagis\ results, except for the missing finite Larmor radius effects in \hagis, leading to higher growth rates of about $+ (\lesssim 1)\%$.\\

\hspace{0.5cm}\begin{minipage}{0.43\textwidth}
  \includegraphics[width=0.98\textwidth, height=4.6cm]{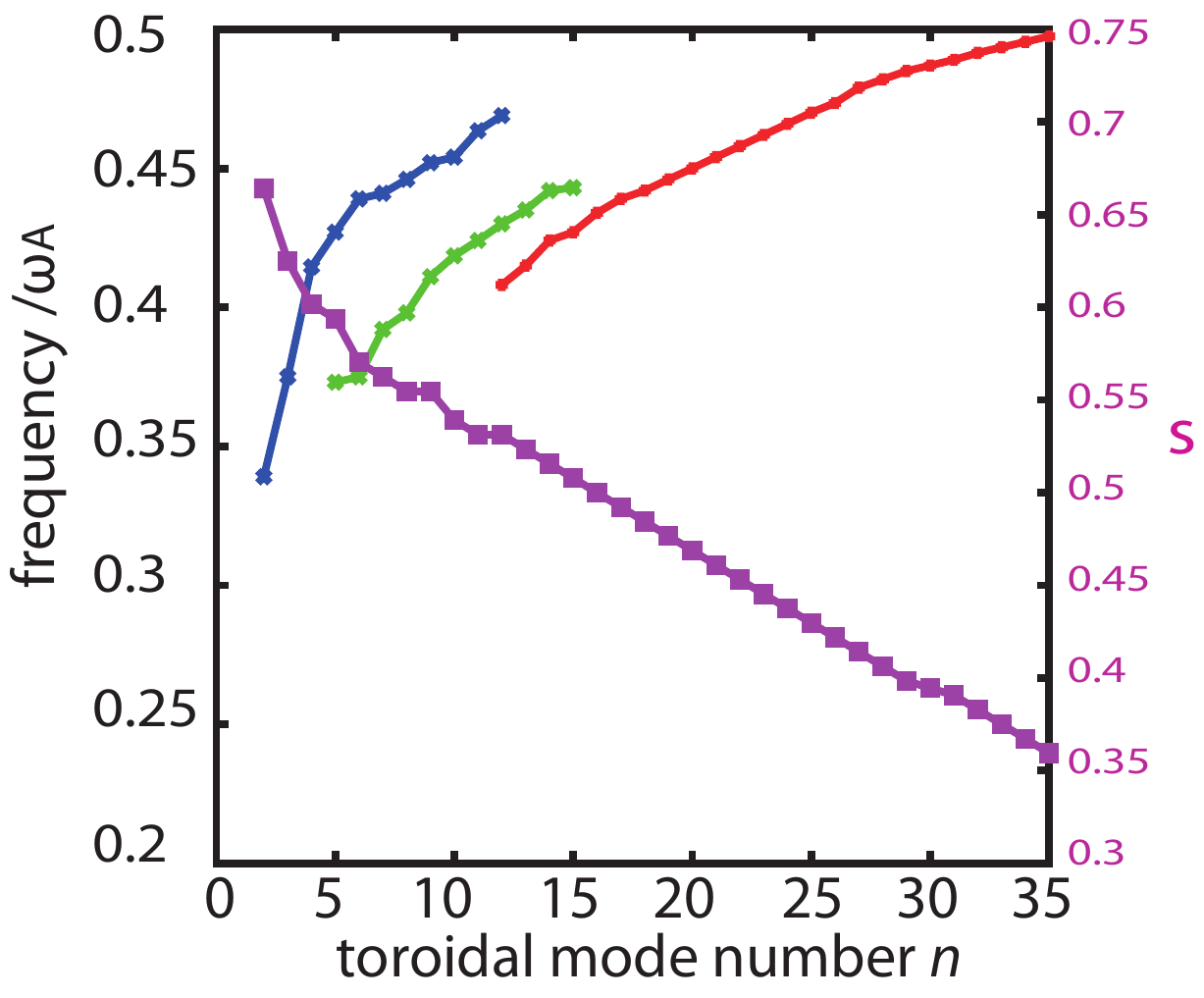}
  \captionof{figure}{\itshape \small \ligka\ calculated TAE frequencies (blue,green and red for low-$n$, intermediate-$n$ and high-$n$ branch) and radial mode position (magenta) for the \iter\ 15 MA scenario with $q_0=0.986$ \cite{LauberVarenna14}.}
  \label{ITER-15MA-q099_freq}
\end{minipage}
\hspace{1cm}\begin{minipage}{0.39\textwidth}
  \includegraphics[width=0.98\textwidth, height=4.6cm]{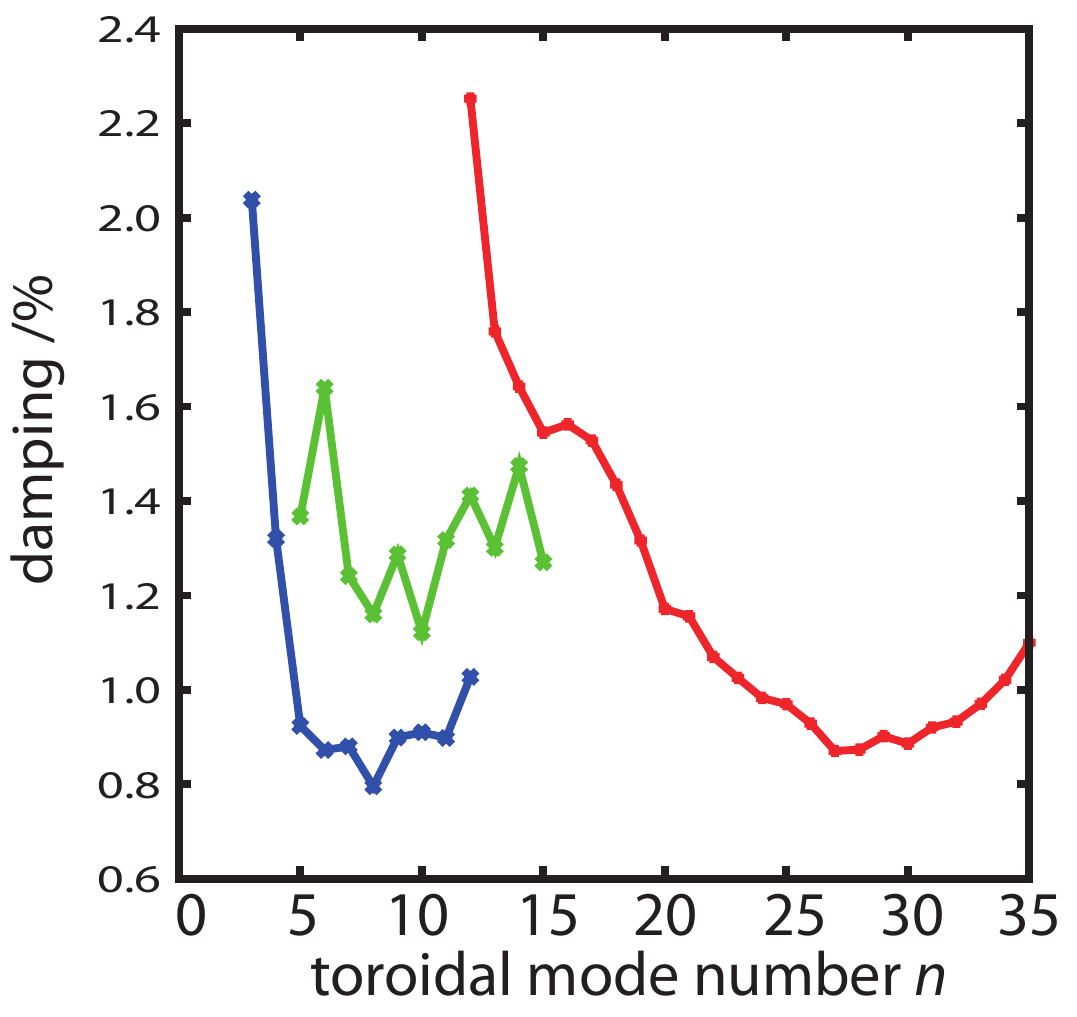}
  \captionof{figure}{\itshape \small \ligka\ calculated TAE damping rates $\gamma_\mr{d}$ (blue,green and red for low-$n$, intermediate-$n$ and high-$n$ branch) for the \iter\ 15MA scenario with $q_0=0.986$ \cite{LauberVarenna14}.}
  \label{ITER-15MA-q099_damp}
\end{minipage}\\

Wave-EP resonance occurs, if the following condition is met: $\omega \approx  -n\langle \dot{\zeta}\rangle + (p \pm m)\langle \dot{\theta}\rangle $, with $\theta$ and $\zeta$ the poloidal and toroidal angle coordinate, $m,n$ the poloidal and toroidal mode number and $p$ the harmonic of the resonance (i.e. $p=0,\pm 1,\pm 2,..$). Due to the flat $q$ profile, the resonance areas in phase space are broad. Where the mode structures cluster radially, the resonances in ($s,E$) space are neighboring, and thus easily overlap for different modes, especially at higher amplitudes. The resonance overlap in velocity space is visualized in\footnote{$\Lambda = \mu B_\mr{mag}/E$ is a constant of motion related to the pitch, with $\mu = E_\perp/B$, the magnetic moment, $B_\mr{mag}$ the equilibrium magnetic field at the axis $s=0$.} \fref{ITER_resplotEL} for three representative modes. Although these modes form different parts of the spectrum, one can clearly see the overlap, especially lower energetic co- and counter-passing, as well as high energetic counter-passing particles.\\
\begin{minipage}{1\textwidth}
  \includegraphics[width=0.48\textwidth, height=5.6cm]{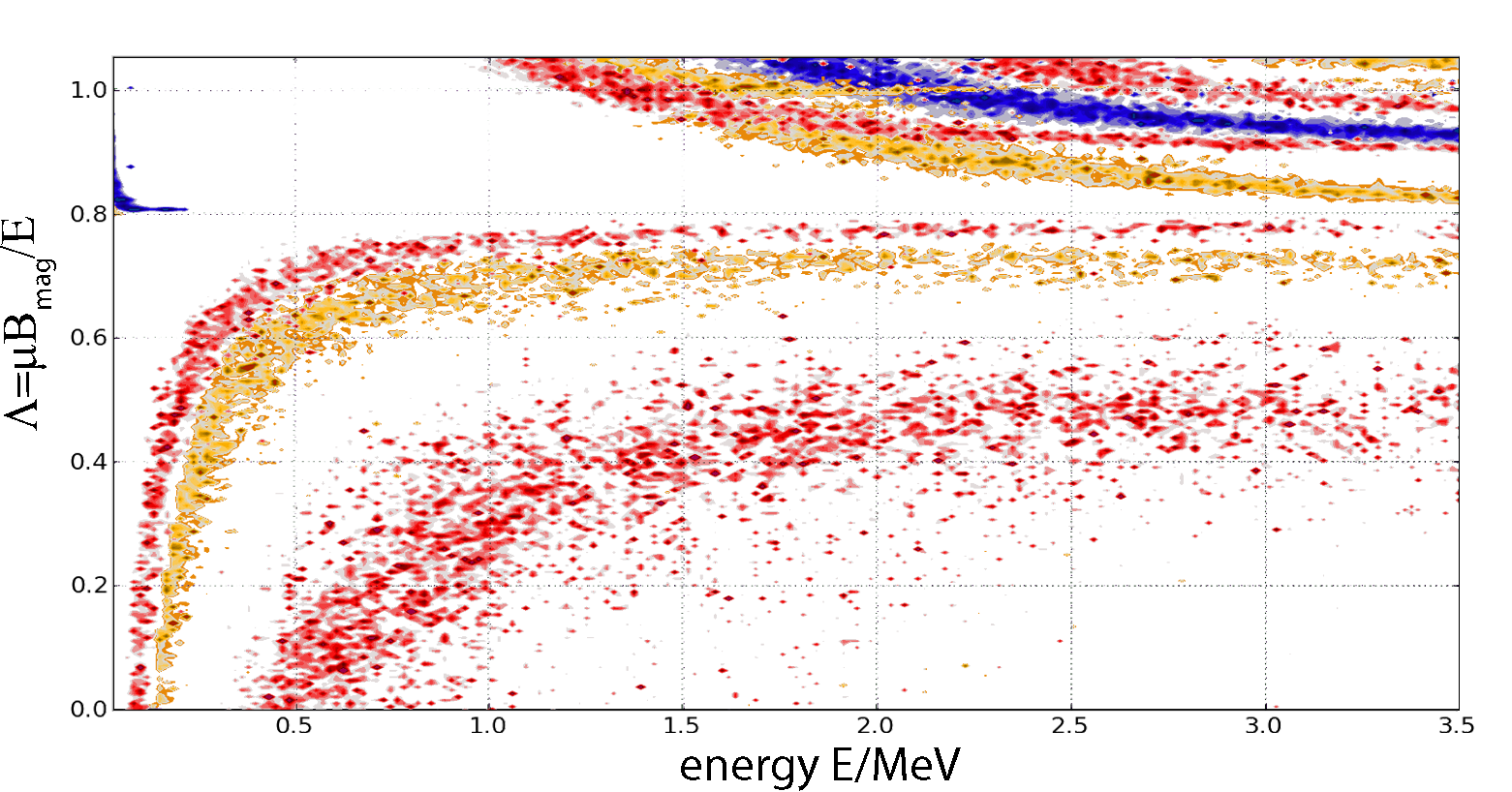}
  \hfill\includegraphics[width=0.48\textwidth, height=5.6cm]{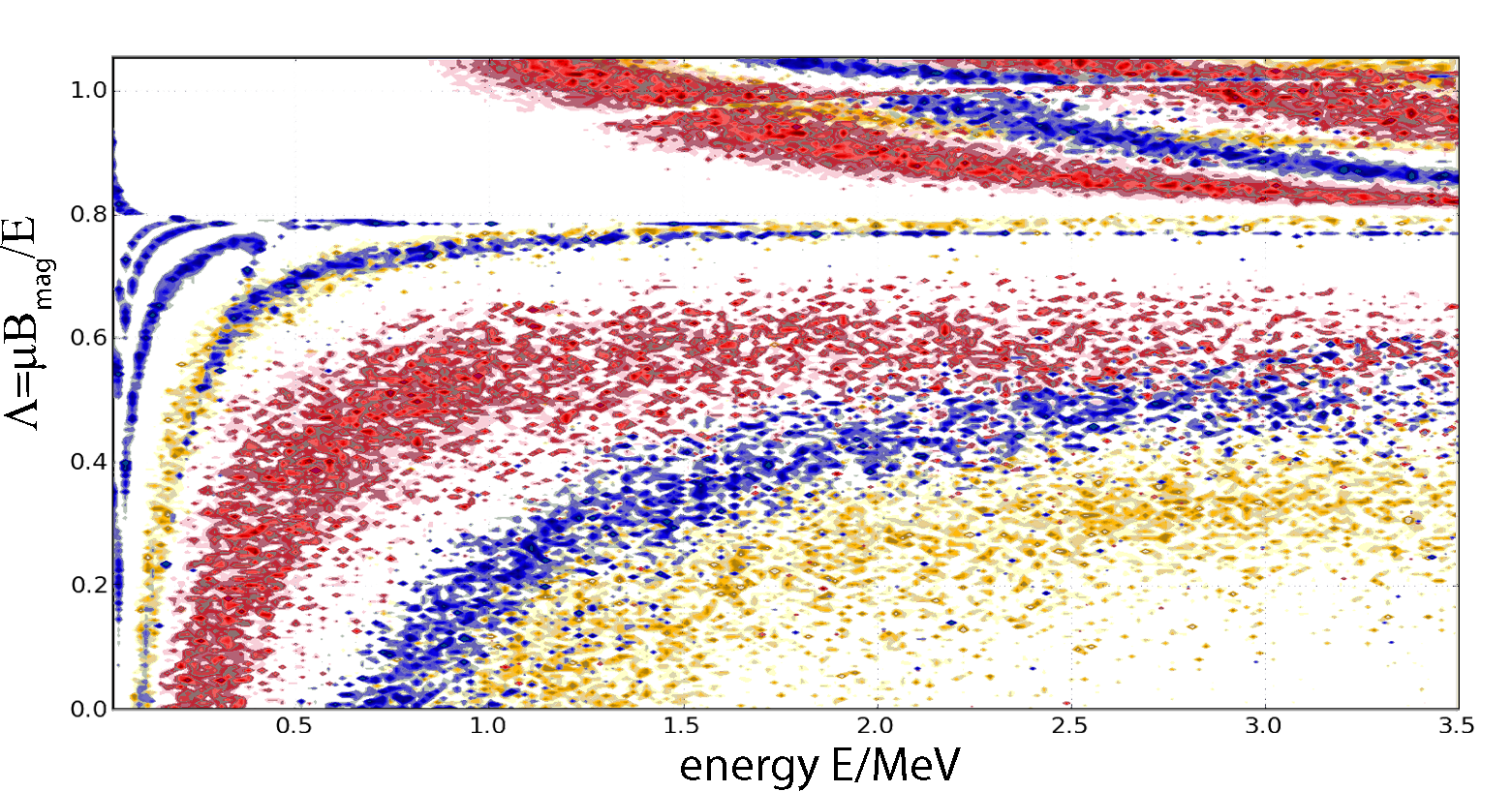}
  \captionof{figure}{\itshape \small Resonances ($p=0,\pm 1$) of representative modes (blue for the low-$n$ branch $n=12,m=14,15,16$ mode, yellow for $n=21,m=21,22$ mode and red for $n=30,m=30,31$ mode) in velocity space ($E,\Lambda$) at radius $s=0.5$, for co- (left) and counter-going (right) particles. For values $\Lambda \gtrsim 0.8$, the particles are trapped.}
  \label{ITER_resplotEL}
\end{minipage}

\section{Modeling Nonlinear Multi Mode EP Transport} \label{sec:Hagisresults}
For the understanding of complex nonlinear multi mode behavior, realistic \iter\ conditions are established step-by-step, which also helps to gain confidence about the numerical conditions. To start with, a simplified \iter\ 15 MA scenario is modeled with \hagis, e.g.\ only selected modes are used instead of all TAEs given by \ligka, and the (small) parallel electric field $E_\|$ is still neglected. Convergence was checked for all numerical parameters at every level of the investigation. In the following, only the most realistic compromise between computational effort and relevant physical features will be discussed. The respective scenario was set up with the least damped 27 TAE modes that were found in the linear analysis in the main branch (red in \fref{ITER-15MA-q099_damp}), $n\in[12,30]$ and in the low-$n$ branch (blue in \fref{ITER-15MA-q099_damp}), $n\in[5,12]$. Since the importance of a detailed poloidal harmonics spectrum is known from earlier \asdex\ Upgrade studies \cite{Schneller13}, all harmonics with a peak greater than 25\% of the mode's maximum peak were taken into account. This affects mostly the low-$n$ branch being simulated with up to 12 poloidal harmonics, whereas the high-$n$ branch is characterized sufficiently by 2 harmonics only. The major motivation to focus on the main and the low-$n$ branch is not only the low damping: together with the effect of multi mode coupling, this can lead -- as will be shown later -- to a nonlinear enhancement of the low-$n$ branch, despite its radial location outside the maximum $\alpha$ particle drive at $s=0.4$. As a consequence, modes with rather high amplitudes can possibly cover a large radial range, and subsequently, EP redistribution, especially in the outer core region can exceed quasi-linear estimates.\\
In order to test the validity of quasi-linear models (which require a reduced computational effort) this section is dedicated to work out a comparison between a quasi-linear (e.g.\ \cite{Bass10,Ghantous12}) and a nonlinear approach. 

\begin{minipage}{0.45\textwidth}
  \includegraphics[width=0.97\textwidth, height=4.3cm]{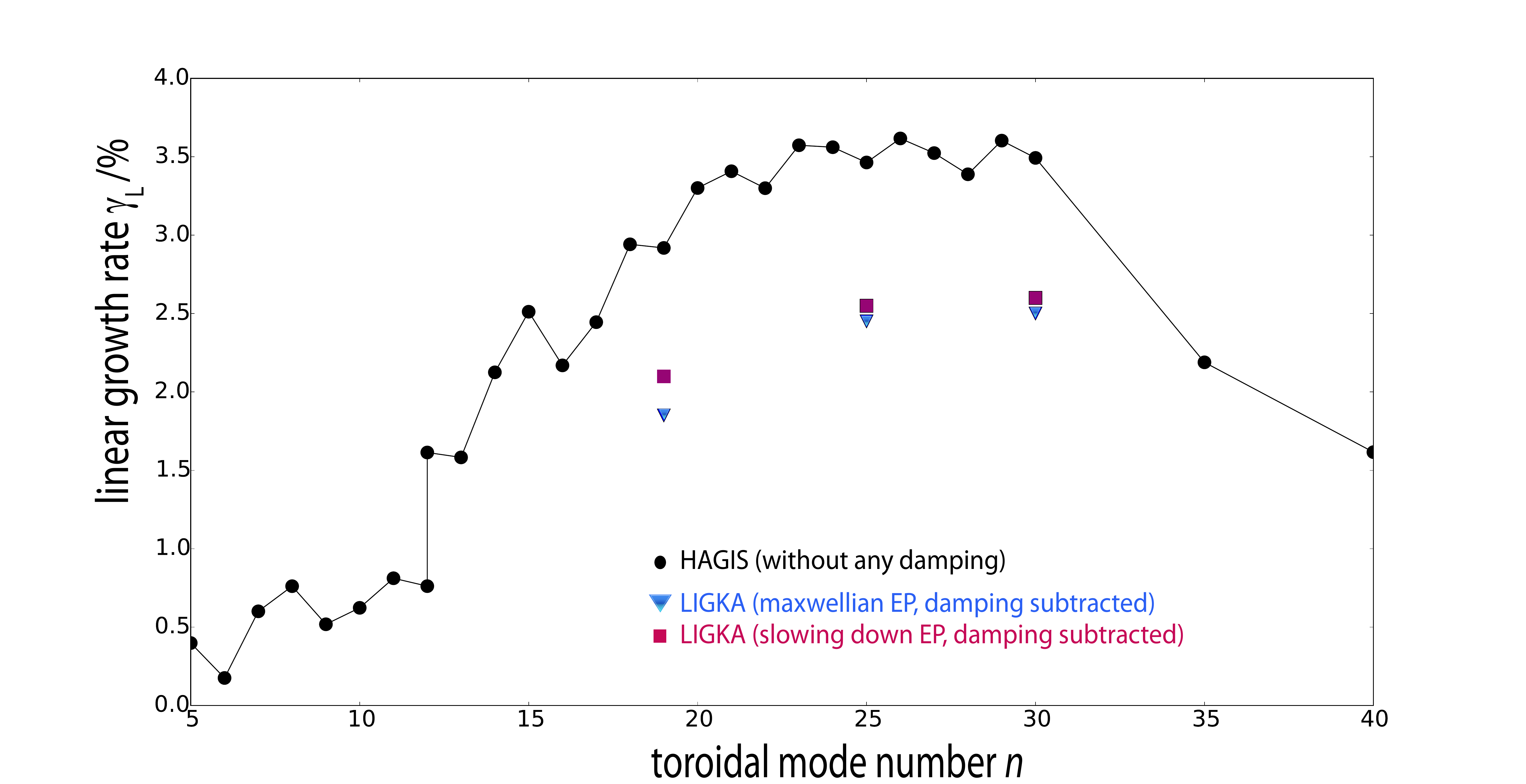}
  \captionof{figure}{\itshape \small Linear growth rates (due to $\alpha$ particle drive) $\gamma_\mr{L}$ (without damping) calculated by \hagis\ (black dots) and \ligka\ (squares and triangles). \hagis\ results are around 1\% higher, due to the missing FLR effects.}
  \label{ITER-gamma_n}
\end{minipage}
\hspace{0.5cm}\begin{minipage}{0.45\textwidth}
  \includegraphics[width=0.97\textwidth, height=4.3cm]{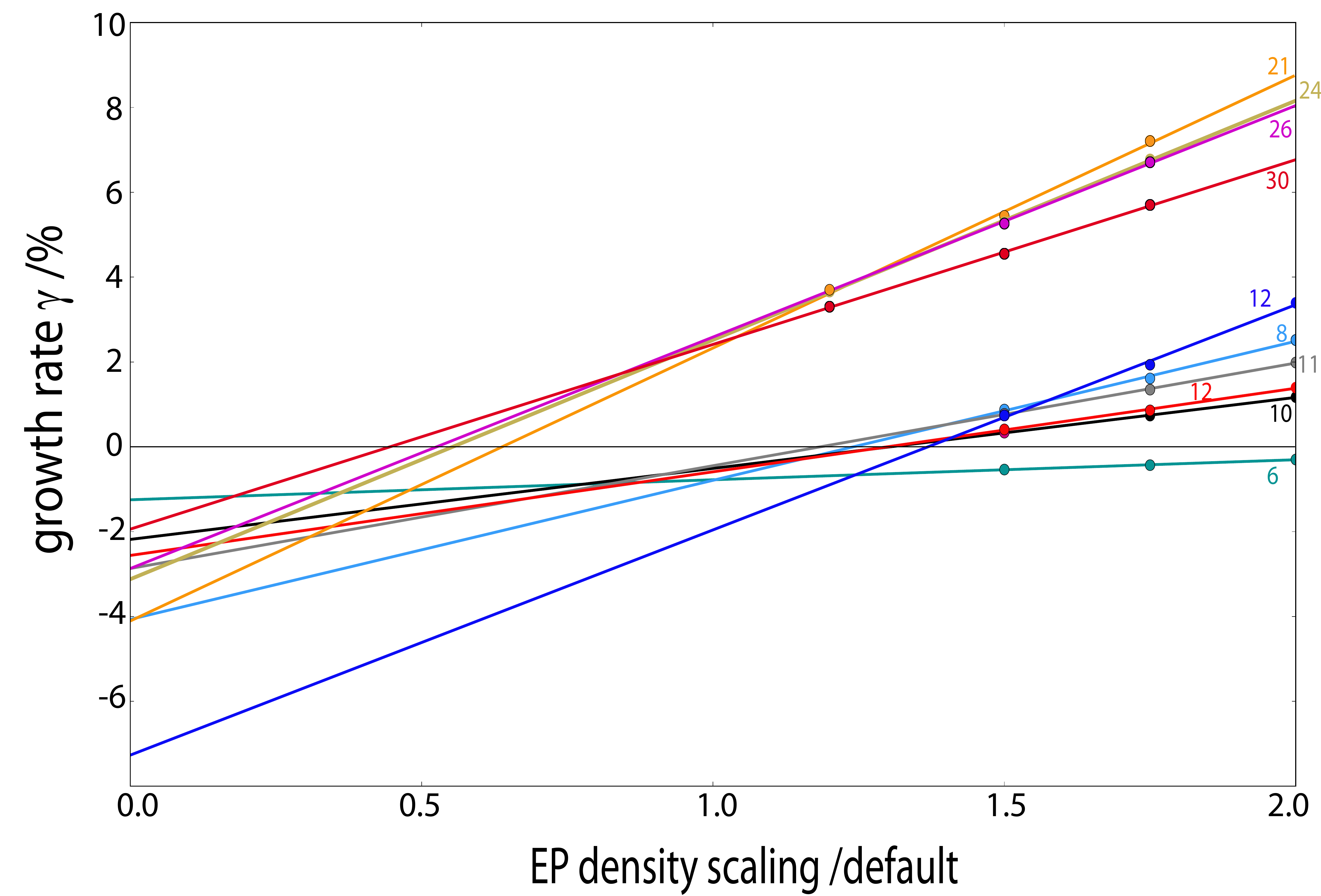}
  \captionof{figure}{\itshape \small Scaling of linear growth rate $\gamma_{\mr{L}}$ subtracted by the \ligka\ damping $\gamma_{\mr{d}}$ over EP density in the single mode cases for selected $n$. Each solid line represents the linear fit for the higher $(\beta,\gamma)$ values of the respective mode $n$ (annotated number), represented by the circles of the respective color.}
  \label{run7singles_densgamma} 
\end{minipage}

In a quasi-linear model, one can estimate the mode growth rates $\gamma$ from linear single mode simulations, since $\gamma \propto \beta_{\mr{EP}}$ (see e.g.\ the\footnote{in the following, $\beta$ denotes the EP plasma beta, i.e.\ ratio of EP pressure over mag.\ background field.} $\beta$ scaling of high-$n$ single mode linear growth rates in \fref{run7singles_densgamma}). Mode saturation amplitudes $A$ can be estimated from quadratical scaling  $A \propto (\gamma/\omega)^2$ w.r.t.\ the single mode linear growth rates over the mode frequency $\omega$. For the \iter\ 15 MA baseline scenario $q=0.986$, the amplitudes in single mode simulations follow quite well such quadratic scaling with the mode linear growth rate, as shown in \fref{run7241-singles_gammaamp}. In general, a critical plasma $\beta_{\mr{crit}}$ can be determined, at which the linear growth rate $\gamma_{\mr{L}}$ equals the damping $\gamma_{\mr{d}}$, and the mode will be marginally stable $\gamma \approx 0$. In quasi-linear theory it is assumed that overlap of resonances causes diffusive EP transport as soon as the linear instability threshold is exceeded, i.e.\ $\beta \geq \beta_{\mr{crit}}$. Subsequently, the diffusive transport relaxes the EP distribution function such that the local $\beta$ takes values around $\beta_{\mr{crit}}$.\\

In the following, a general comparison between nonlinear and quasi-linear model is carried out. The main purpose of \sref{sec:nonlin_wodamp} and \sref{sec:nonlin_ligkadamp} is to point out limits beyond which enhanced EP transport can happen, triggered by a domino effect. These limits concern damping as well as EP density -- however, a thorough study scanning both parameters goes beyond the scope of this work. Instead, \sref{sec:nonlin_wodamp} shows crossing the limit to trigger domino-like EP transport via reducing the damping, whereas \sref{sec:nonlin_ligkadamp} explores the limit with respect to the EP density.

\subsection{Nonlinear v.s.\ Quasi-linear Transport under Reduced Damping}\label{sec:nonlin_wodamp}
A general comparison between nonlinear and quasi-linear model is carried out with reduced mode damping. The damping is reduced by around a factor of 6, which is in a way, that domino-like EP behaviour can be observed nonlinearly, but critical gradients are still determinable. Further advantages are hereby the reduced computational effort and the fact, that a possible difference between quasi-linear and nonlinear model can be shown very clearly.\\
\Fref{run7206_betacrit-100} shows the final EP density gradient depletion caused by all 27 relevant modes together, simulated with amplitudes fixed at their single mode (peak) saturation level (in the following, this model will be referred to as ``quasi-linear'' \hagis\ simulation). The derivative of it is comparable but slightly above the local gradient of $\beta_{\mr{crit}}$ at each mode location (thicker lines). Thus, $\beta_{\mr{crit}}$ provides a lower estimate for profile relaxation here.\\
In the following, it will be investigated, whether this quasi-linear estimate is a valid assumption by modeling the mode evolution of all relevant 27 modes self-consistently. For a radial EP density profile scaled to 30\% of the default value, the quasi-linear estimate predicts all modes stable, except $n=29$. The self-consistent nonlinear multi mode simulation reveals no mode growth for the low-$n$ branch and very moderate growth ($\gamma < 0.8\%$) for the high-$n$ branch. For a radial EP density profile scaled to 50\% of the default value, the quasi-linear estimate predicts mode growth for $n \in [23,30]$. For this case, the quasi-linear approach is significantly underestimating the nonlinear multi mode amplitude evolution: the modes $n \in [23,30]$ exceed the estimate (by a factor of up to 28). Also, the modes  $n \in [12,22]$ grow moderately. Although the quasi-linear prediction underestimates the nonlinear amplitude evolution, none of the modes exceed $\delta B/B_\mr{mag} \approx 6\cdot 10^{-4}$, and the low-$n$ branch remains marginally unstable. Thus, no significant EP redistribution is found, in agreement with the quasi-linear approach.\\

\hfill\begin{minipage}{0.35\textwidth}
  \includegraphics[width=0.99\textwidth]{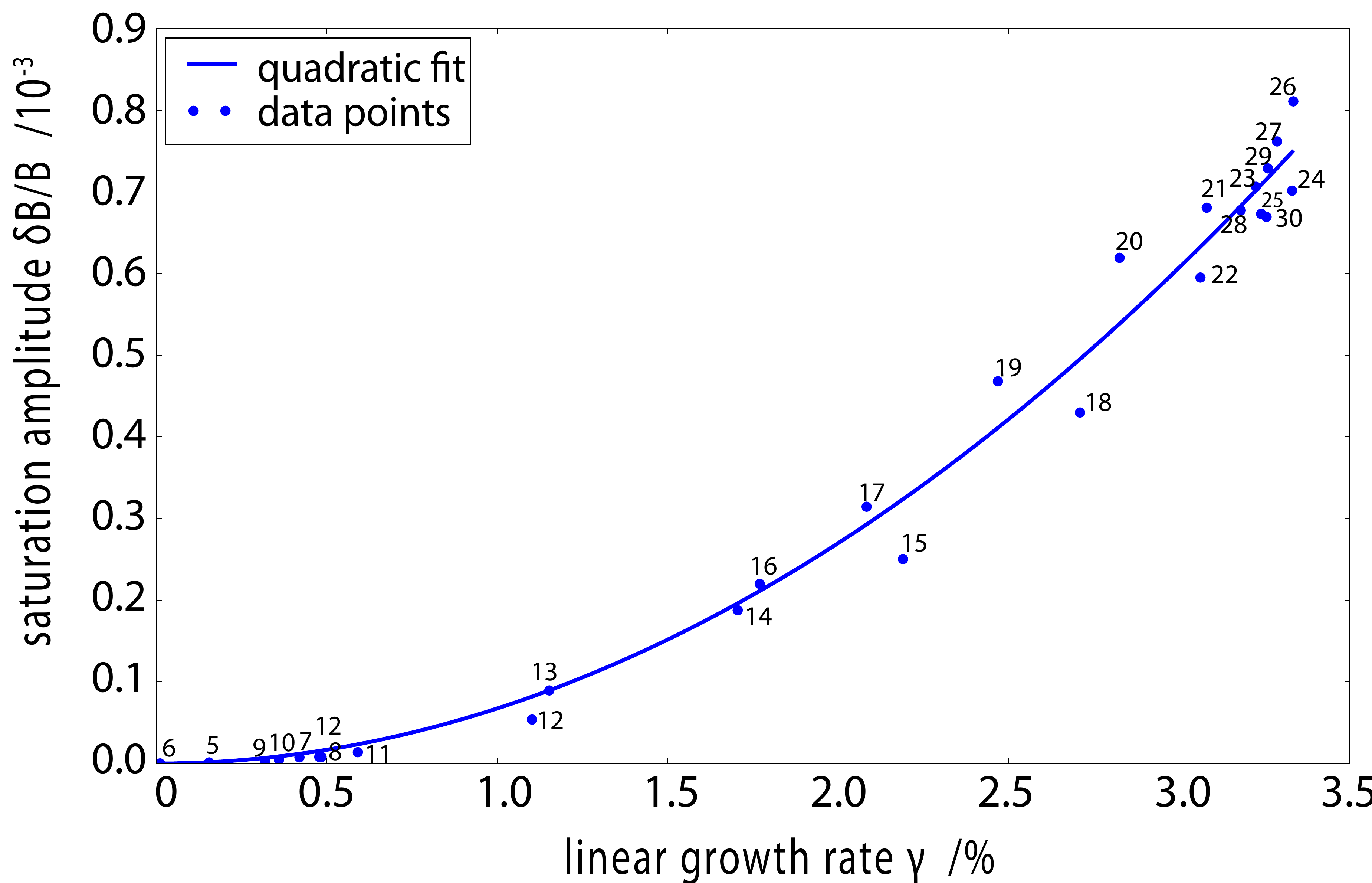}
  \captionof{figure}{\itshape \small Saturation amplitude depending on growth rate for single mode simulations at the default radial EP density profile. The annotated labels give the toroidal mode number $n$.}
  \label{run7241-singles_gammaamp}
\end{minipage}
\hfill\begin{minipage}{0.6\textwidth}
  \includegraphics[width=0.95\textwidth]{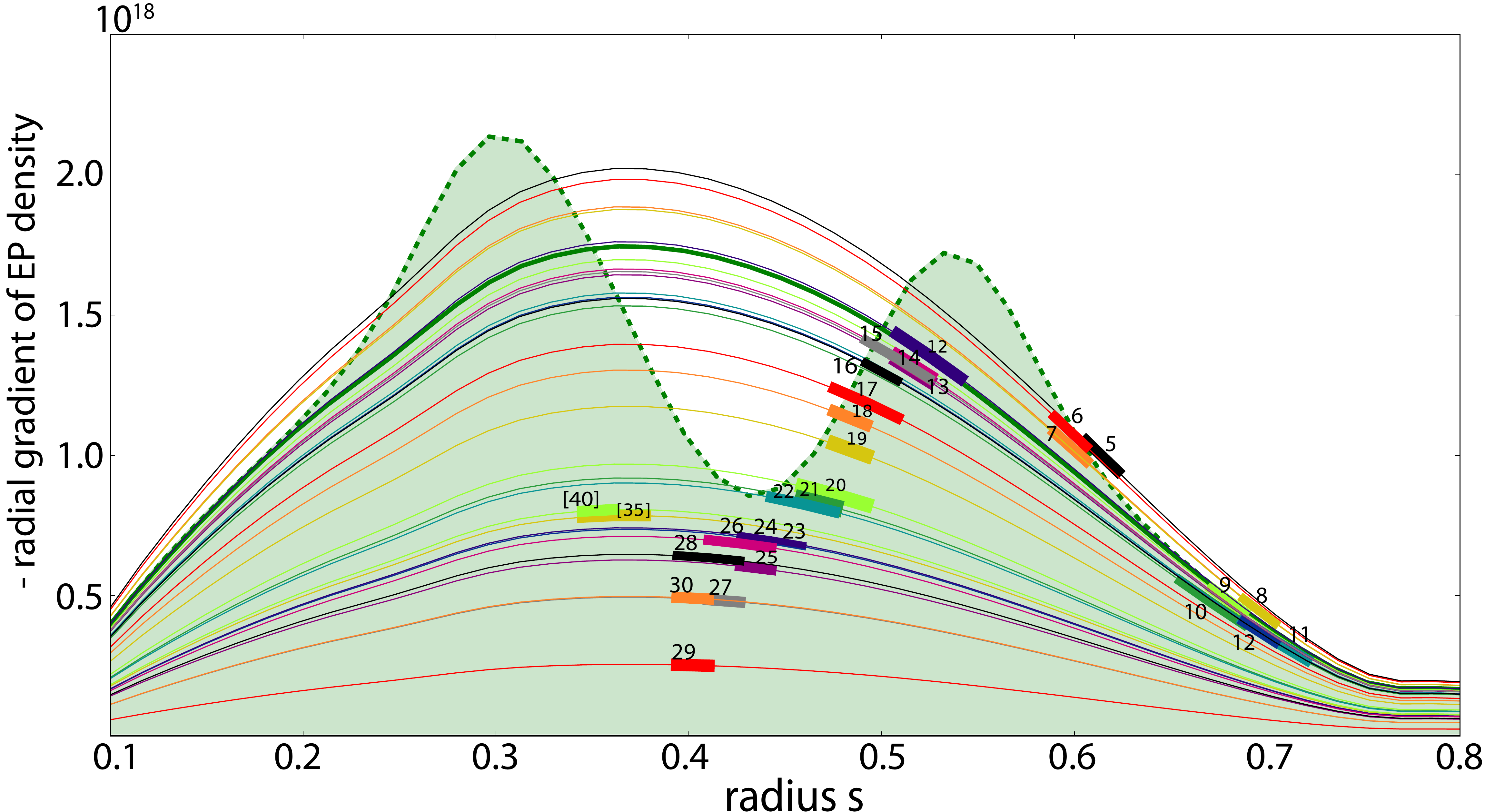}
  \captionof{figure}{\itshape \small ``Quasi-linear'' radial EP density gradient depletion: multi mode simulation with amplitudes fixed to single mode saturation levels (under reduced damping). The thick solid green line shows the initial, the dashed green line (with shadowed area beneath) the final value of radial density gradient. The colored lines indicate the gradient of the critical EP density for every mode $n$ (under reduced damping). The thicker section of the lines visualize the radial position of the mode structure peaks.}
  \label{run7206_betacrit-100}
\end{minipage}\\

If the radial EP density profile is scaled up to its default value as predicted for this \iter\ scenario, the self-consistent wave evolution in the multi mode scenario differs significantly from the single mode simulation, i.e.\ from the quasi-linear expectation: at around $3~\mr{ms}$, the waves of the low-$n$ branch (blue in \fref{run7241+singles_ampl}) get strongly excited and their amplitude grows to values of roughly more than one order of magnitude (a factor of 5 to 65) higher than in the single mode saturation (light blue in \fref{run7241+singles_ampl}). While nonlinearly, some of the low-$n$ modes (especially $n=12,11$) become one of the dominant modes, the ratio of linear growths $\gamma/\omega$ in the multi mode over the single mode cases is similar ($1\pm0.1$) for the modes $n>12$ (for the $n \leq 12$ modes it is comparable ($1\pm0.6$) in the early\footnote{in the later linear phase of multi mode simulations, the linear growth rate is not constant anymore for the low-$n$ modes.} linear phase). This emphasizes the strong nonlinear effect of the multi mode behaviour, that cannot be foreseen linearly.\\
The amplitudes of the high-$n$ branch modes (red in \fref{run7241+singles_ampl}) in the multi mode case reach values that are enhanced slightly, by a factor of 5 at most for $n>12$. For both, low- and high-$n$ branch the nonlinear multi mode dynamics is non-trivial -- modes reach high amplitude regimes at very different times $t$. While the low-$n$ branch grows in 3 phases ($t < 0.4~\mr{ms}$: like single mode, $0.4 ~\mr{ms} < t < 3.0~\mr{ms}$: first enhancement until a slight ``saturation'', $t > 3.0~\mr{ms}$: second enhancement), the behavior of the high-$n$ branch starts to differ from single mode evolution only at the onset of its saturation ($t\approx 0.6~\mr{ms}$). This saturation is higher than in the single mode case and followed by a second enhancement of high-$n$ modes at around $t\approx 4.5~\mr{ms}$, when the low-$n$ modes have reached relevant amplitudes of $\delta B/B_\mr{mag} \approx 10^{-3}$.\\

To understand the nonlinear multi mode behaviour, it is helpful to look at the temporal evolution of EP redistribution: until $t\approx 4~\mr{ms}$, the radial redistribution does not exceed the relaxed profile observed in the fixed amplitude multi mode simulation (green lines in \fref{run7206+237+241_disbs}). However, this redistribution not only leads to a steeper EP density gradient at the radial position of the low-$n$ modes, but also provides more EPs at the radial location of these modes.  There is a critical redistribution that triggers the excitation of the low-$n$ modes, which can be reached only due to the large amount of more core-localized modes in the high-$n$ branch: a reduced scenario simulation of only the highest amplitudes and the radially most extended modes ($n=8,11,12,12,18,21,24,30$) did not lead to sufficient redistribution to excite the low-$n$ modes, although the high-$n$ modes reached similar saturation amplitude levels compared to the full scenario simulation. Only with the low-$n$ modes reaching levels around $\delta B/B_\mr{mag} \approx 10^{-3}$, massive EP gradient depletion sets in, and EPs are redistributed radially outwards to $s\approx 0.8$ (see blue line in \fref{run7206+237+241_disbs}). A similar depletion occurs even if the already quasi-linearly relaxed EP density profile was chosen as initial condition (see red line in \fref{run7206+237+241_disbs}). This is remarkable, since it may indicate a possible overshoot effect, leading to domino-like behaviour from an initially quasi-linear state.\\
\begin{minipage}{0.59\textwidth}
  \includegraphics[width=0.99\textwidth]{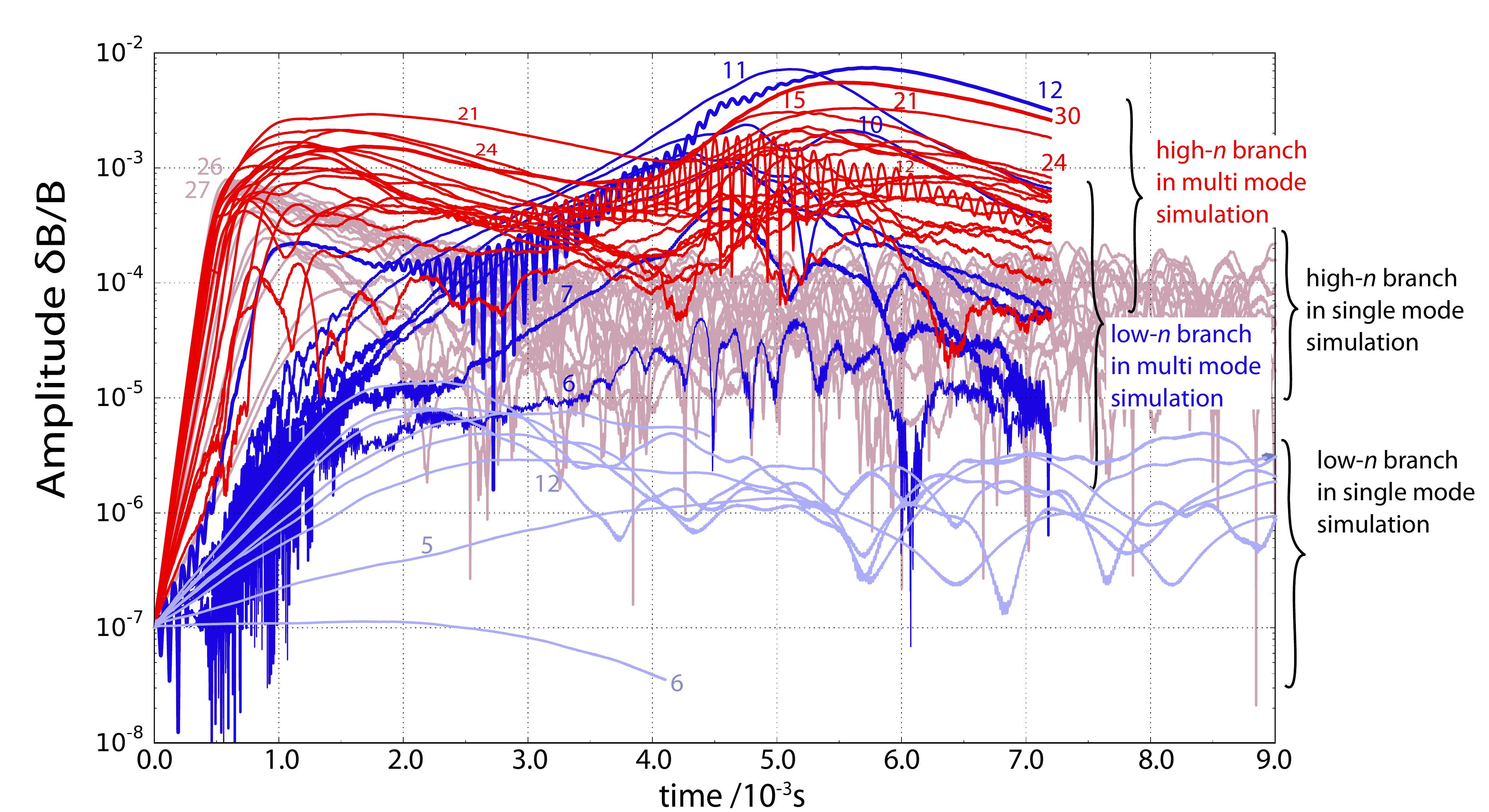}
  \captionof{figure}{\itshape \small Default radial EP density profile case with reduced damping: self-consistent multi mode evolution (strong colored solid lines) compared to the single mode evolution for all 27 modes (light colored solid lines). Blue represents modes of the low-$n$ branch, red of the high-$n$ branch. The integers denote toroidal mode numbers $n$.}
  \label{run7241+singles_ampl}
\end{minipage}
\hfill\begin{minipage}{0.40\textwidth}
  \includegraphics[width=0.88\textwidth]{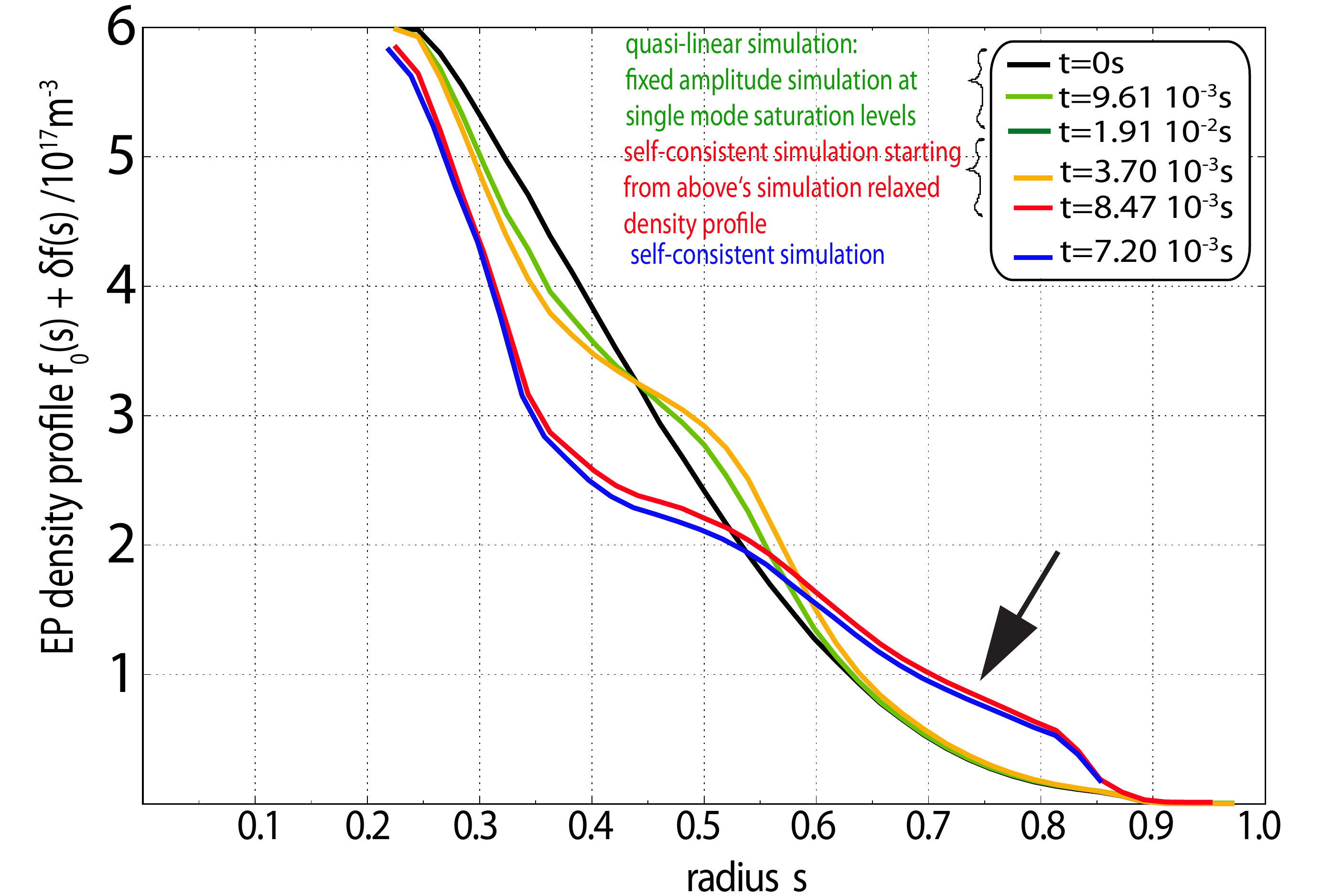}
  \captionof{figure}{\itshape \small Radial EP redistribution for the self-consistent multi mode simulation with reduced damping (orange line for an earlier time point, red line for the final time point) starting from the quasi-linearly relaxed EP density profile (converged over time to the dark green line). The blue line is the final EP distribution in the self-consistent multi mode simulation starting from the default initial EP density profile (black line).}
  \label{run7206+237+241_disbs}
\end{minipage}\\

In the evolution of the density gradient depletion, big differences can be observed between the self-consistent nonlinear multi mode scenario (with 27 modes) and the quasi-linear \hagis\ simulation with the same modes (amplitudes fixed at the respective single mode saturation level): while the quasi-linear case converges in time towards a depletion slightly above the local values of $\nabla \beta_{\mr{crit}}$ (\fref{run7206_betacrit-100}), the nonlinear scenario also reaches this state (\fref{run7241+206_betacrit}, left), but with the low-$n$ modes exceeding amplitude $\delta B/B_\mr{mag}$ levels of $10^{-3}$, broad redistribution sets in. The density gradient in the outer core region is depleted rapidly (\fref{run7241+206_betacrit}, right), triggered by the radial overlap of the growing low-$n$ modes with the higher-$n$ modes (this is visualized by the thick lines in \fref{run7241+206_betacrit}, left, while the necessary overlap in velocity space was shown in \fref{ITER_resplotEL}). This domino behaviour is clearly a non-local effect, which is avoided in the quasi-linear scenario by a transport barrier between the outer, low-$n$ and the inner, high-$n$ branch: the locally very high EP density gradient (\fref{run7241+206_betacrit}, left) around $s\approx 0.55$ cannot be depleted because there is no radial mode overlap at amplitudes above certain threshold (around $\delta B/B_\mr{mag} \approx 10^{-3}$). The situation changes rapidly, as soon as broad, low-$n$ modes grow, triggered by gradient steepening due to EP redistribution from the inner, high-$n$ modes.\\

\begin{minipage}{0.88\textwidth}
  \centering
  \includegraphics[width=0.90\textwidth]{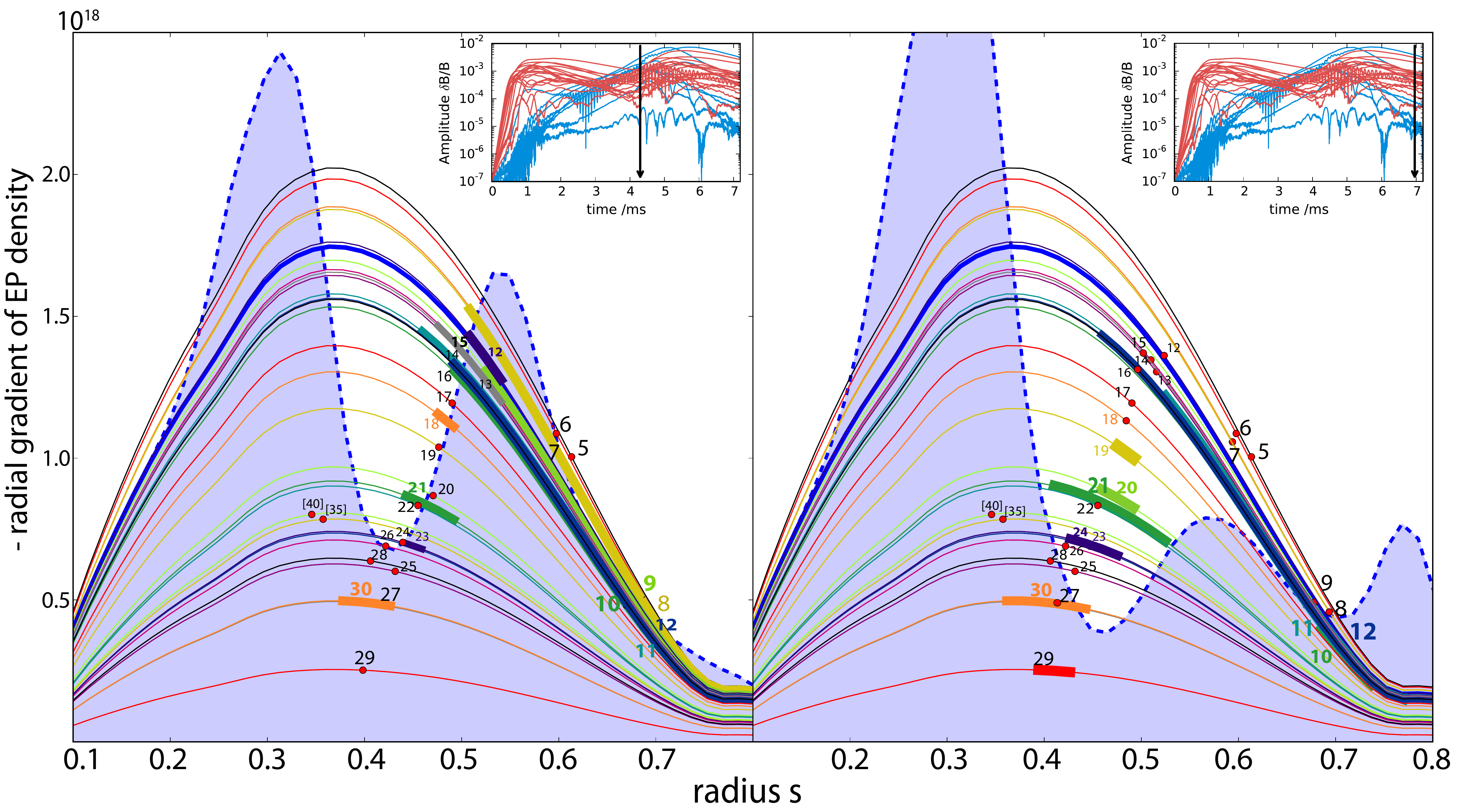}
  \captionof{figure}{\itshape \small Nonlinear radial EP density gradient depletion in the self-consistent multi mode simulation (27 most relevant modes) with default initial EP density profile (thick blue line) under reduced damping, for two different times during the simulation (left: $t=4.3~\mr{ms}$, right: $t=6.9~\mr{ms}$, see arrow in the respective insert for mode amplitude constellation at that time). The thick solid blue line shows the initial value of radial density gradient, the dashed blue line (with shadowed area beneath) its value at time $t$. The colored lines indicate the gradient of the critical EP density of each mode $n$. The thicker section of the lines visualize the radial extent, where the respective mode amplitude $\delta B/B_\mr{mag}$ exceeds $10^{-3}$ at time $t$.}
  \label{run7241+206_betacrit}
\end{minipage}

\subsection{Nonlinear v.s.\ Quasi-linear Transport with Damping given by LIGKA}\label{sec:nonlin_ligkadamp}
In this section, the original damping is used as given by the \ligka\ code. With the default EP density profile, the low-$n$ branch stays now at low saturation amplitudes, well separated from the high-$n$ branch (\fref{run7254_ampl}). As a consequence, there is no significant EP redistribution from the low-$n$ branch. Due to this fact and the stronger damping, the high-$n$ branch does not reach the level of massive redistribution either. Thus, no domino-like interaction similar to the situation described in the last section takes place.\\

On the way towards a more realistic \iter\ prediction, there are still missing effects to be included into the model. The most important one is the destabilization caused by the NBI generated EP distribution, as shown in \fref{ITER_alpha+beamdrive}: in Ref.\ \cite{Pinches15}, it is shown that the NBI drive can -- depending on the radial position and the mode number -- more than double the drive for the here investigated \iter\ scenario. To include this effect, a multi-species model is needed, which has been implemented into the \hagis\ code and is now in the testing phase. Results will be shown in a follow-up publication. In this article, the nonlinear EP transport investigation is restricted to the single-species model, using just $\alpha$ particle drive. To investigate the possibility of a domino-effect under stronger drive, the $\alpha$ particle density is scaled to 200\%. The rest of this section will be dedicated to an investigation of the multi mode behaviour under such scaling.\\
\begin{minipage}{0.59\textwidth}
  \includegraphics[width=0.9\textwidth]{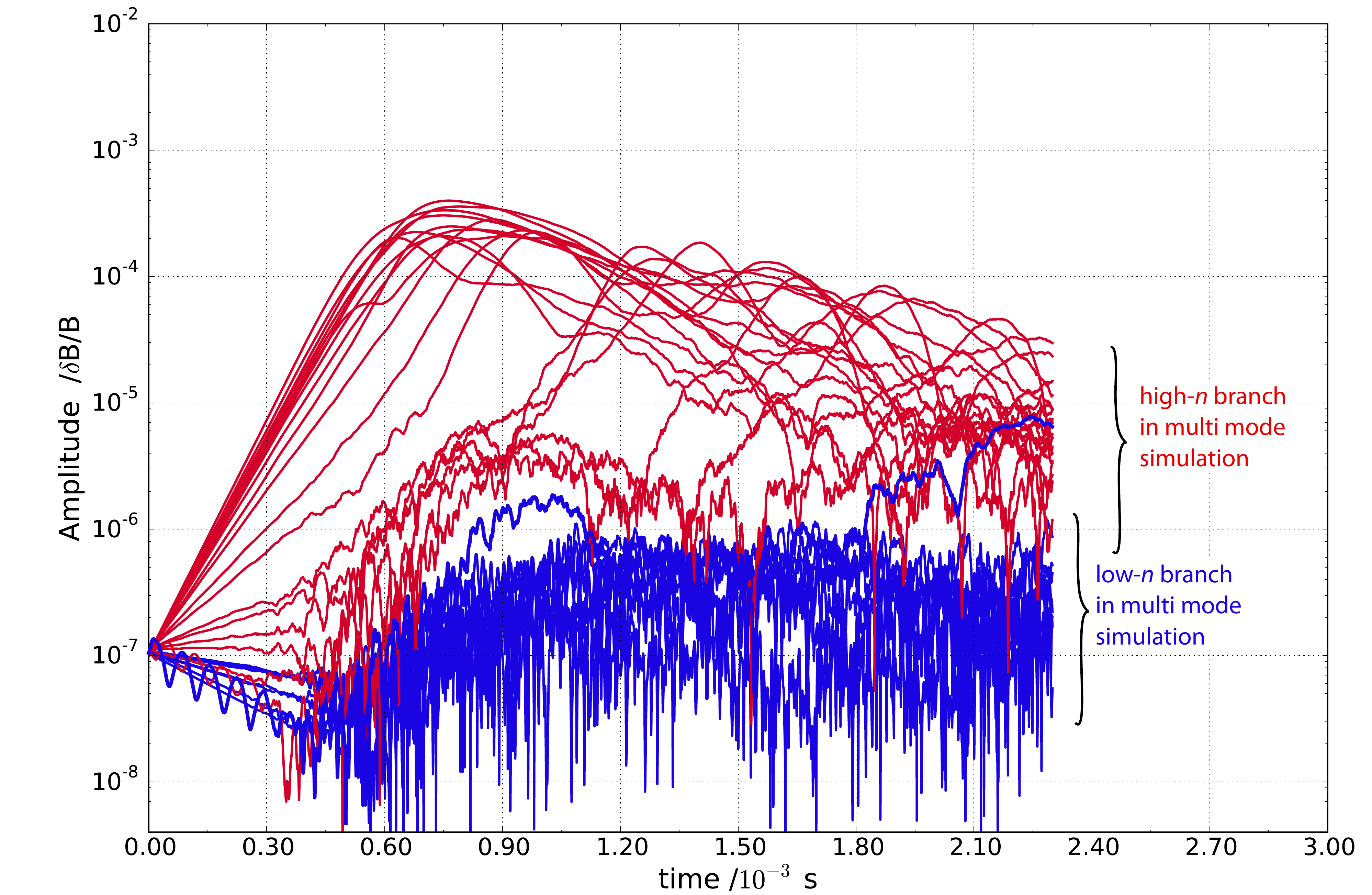}
  \captionof{figure}{\itshape \small Default radial EP density profile case: self-consistent multi mode evolution for all 27 modes. Blue represents modes of the low-$n$ branch, red of the high-$n$ branch. The integers denote toroidal mode numbers $n$.}
  \label{run7254_ampl}
\end{minipage}
\begin{minipage}{0.39\textwidth}
  \includegraphics[width=1\textwidth]{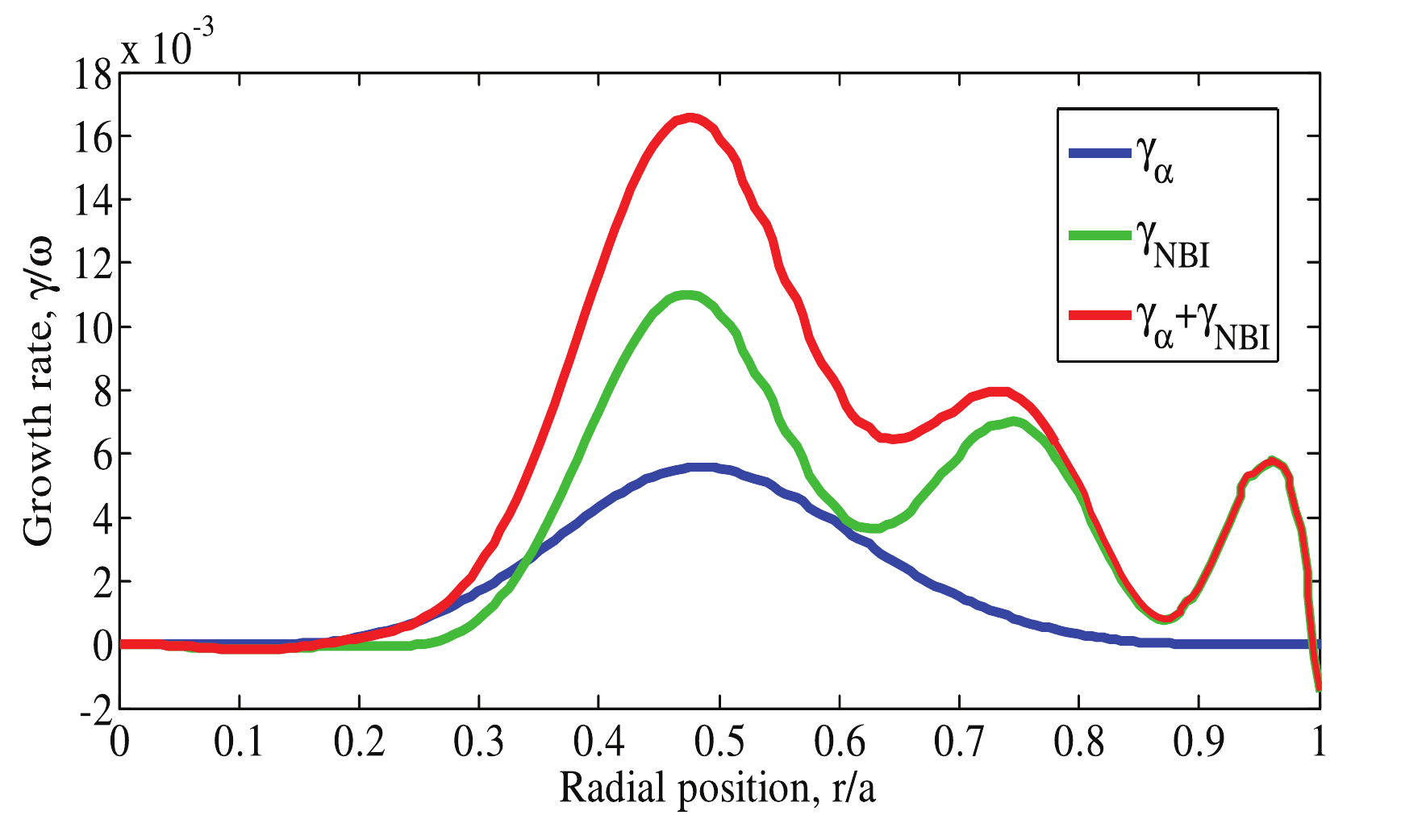}
  \captionof{figure}{\itshape \small Radial profile of the $\alpha$ particle (green) and NBI (blue) drive, and both of them together (red). From \cite{Pinches15}.}
  \label{ITER_alpha+beamdrive}
  \vspace{3cm}
\end{minipage}\\
\begin{minipage}{0.59\textwidth}
  \includegraphics[width=0.90\textwidth]{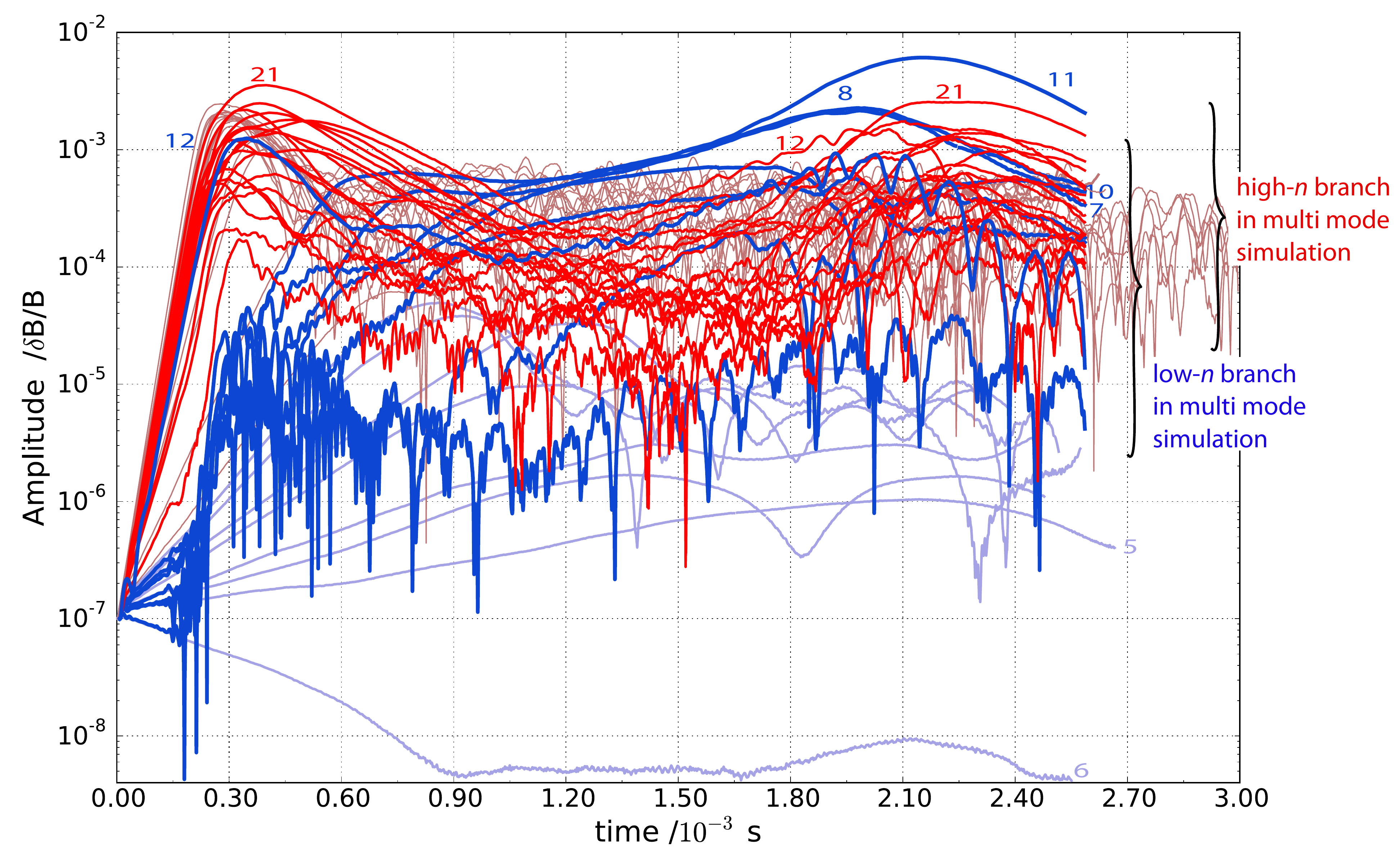}
  \captionof{figure}{\itshape \small Case of EP density profile scaled to 200\%: self-consistent multi mode evolution (strong colored solid lines) compared to the single mode evolution for all 27 modes (light colored solid lines). Blue represents modes of the low-$n$ branch, red of the high-$n$ branch. The integers denote toroidal mode numbers $n$.}
  \label{run7258_ampl}
\end{minipage}
\begin{minipage}{0.40\textwidth}
  \includegraphics[width=0.99\textwidth]{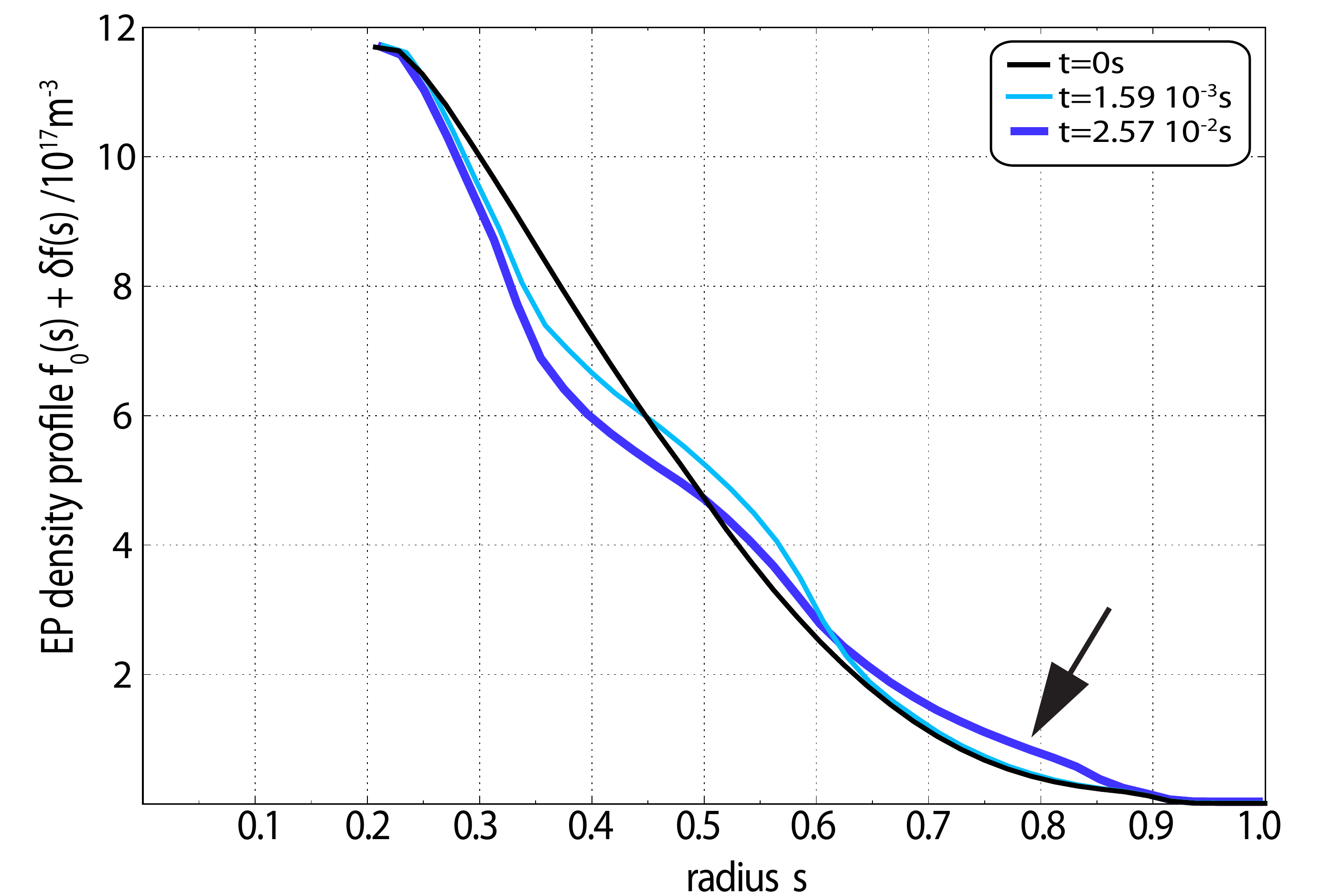}
  \captionof{figure}{\itshape \small Radial EP redistribution for the self-consistent multi mode simulation (light blue line for an earlier time point, dark blue line for the final time point) starting from the 200\% initial EP density profile (black line).}
  \label{run7258_disbs}
  \vspace{1.5cm}
\end{minipage}\\
In this case, as shown in \fref{run7258_ampl}, a domino-like behaviour is visible, enabling the low-$n$ modes to reach higher amplitudes compared to the single mode case -- some ($n=11,8,10$) reach $\delta B/B_\mr{mag} > 10^{-3}$ and become dominant in the nonlinear phase. As a consequence, a second enhancement of the high-$n$ mode saturation is triggered. However, with the higher drive and at the same time higher damping, the temporal behaviour of the mode amplitude evolution relative to each other is very different compared to the results of the last section: the high-$n$ modes grow very fast and have already started decaying by the time the highly damped low-$n$ modes reach the levels of significant redistribution. Therefore, the peak amplitudes of many of the high-$n$ modes are reached at the very beginning of the nonlinear phase, while the second peak is lower (up to a factor of $\lesssim 6$). As a consequence, the depletion of the EP density gradient in the region around $s \approx 0.6$ is weaker (see \fref{run7289+258_betacrit-095}, right). Due to the high amplitudes of certain low-$n$ modes (mainly $n=11$). However, the previously observed outer EP redistribution up to $s \approx 0.8$ is seen here as well (although weaker), and cannot be found in the quasi-linear estimate (\fref{run7289+258_betacrit-095}, left), although the quasi-linear gradient depletion in the area $s \approx 0.45$ is stronger than reported in the previous section.

 \section{Conclusions and Discussion}
 This work presented a nonlinear investigation on the interaction of energetic particles with multiple global modes in the \iter\ 15 MA baseline scenario using the hybrid \hagis-\ligka\ code package. The addressed question is, if the overall energetic particle transport remains within the quasi-linear estimates or if possible nonlinear excitation of linearly stable TAEs via phase space coupling effects leads to enhanced, domino-like energetic particle transport. The challenge for an investigation arises not only from the high amount of modes and poloidal harmonics, but also from the high resolution range needed due to the large machine size combined with high toroidal mode numbers $n$. For the \iter\ 15 MA scenario with $q_0 = 0.986$ the linear, gyrokinetic, non-perturbative code \ligka\ predicts a radially dense cluster of TAEs up to toroidal mode number $n \approx 40$ that can be categorized into three different branches: low-$n$, intermediate-$n$ and high-$n$. As discussed in ref.\ \cite{LauberVarenna14} a flat $q$ with $q_0$ close to 1 causes a dense cluster of modes that facilitates the excitation of edge TAEs. Therefore, the $q_0 = 0.986$ scenario investigated in this work is a ``worst-case'', compared to other scenarios with different $q$ shaping (as provided in ref.\ \cite{Pinches15}).\\
\begin{minipage}{0.88\textwidth}
  \centering
  \includegraphics[width=0.9\textwidth]{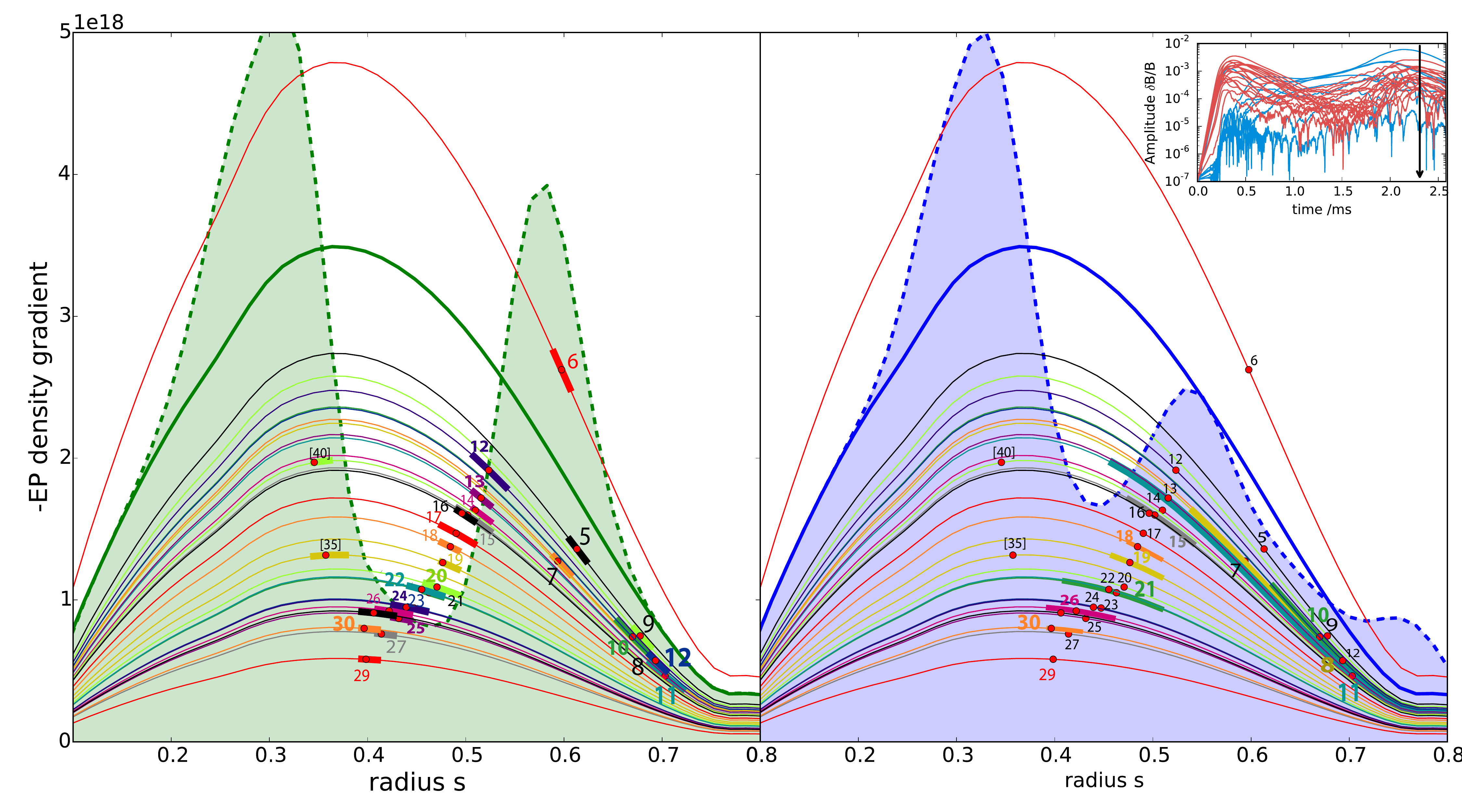}
  \captionof{figure}{\itshape \small Radial EP density gradient depletion in the quasi-linear (left) and the nonlinear self-consistent multi mode simulation (right) with 27 most relevant modes at 200\% initial EP density profile (thick green/blue line) at $t=2.45~\mr{ms}$ during the simulation (see arrow in the insert for mode amplitude constellation at that time). The thick solid green/blue line shows the initial value of radial density gradient, the dashed green/blue line (with shadowed area beneath) its value at time $t$. The colored lines indicate the gradient of the critical EP density of each mode $n$. The thicker section of the lines visualize the radial extent, where the respective mode amplitude $\delta B/B_\mr{mag}$ exceeds $10^{-3}$ at time $t$.}
  \label{run7289+258_betacrit-095}
\end{minipage}\\

Nonlinear \hagis\ simulations were carried out, taking into account the least damped modes of the core-localized high-$n$ branch ($n\in[12,30]$) and the weakly damped low-$n$ branch ($n\in[5,12]$) with modes in the outer core region. It was shown that in our model, that the EP redistribution stays well within quasi-linear expectations -- even if damping is neglected -- if a certain EP density threshold is not overcome: the relaxed radial density gradient remains well above the critical gradient. In the scenario with full, default EP density profile and neglected damping, it was found, that the phase space redistribution caused by the (sufficiently large) number of inner, high-$n$ modes triggers the excitation of the otherwise marginally unstable outer modes of the low-$n$ branch (enhancement of a factor between 5 and 60). This excitation is very sensitive to the amount of redistribution by the high-$n$ modes. It leads to the rapid depletion of the radial EP density gradient in the region of $s \approx 0.55$. This non-local effect is based on the radial overlap of the two branches that forms, at the time when the radially broad low-$n$ mode amplitudes grow towards $\delta B/B_\mr{mag} \approx 10^{-3}$. Therefore, such depletion does not occur in the quasi-linear scenario, with mode amplitudes fixed to the much lower single mode saturation levels, which effectively also restricts the modes locally and therefore creates a transport barrier. The same transport barrier is present for more favorable $q$ profiles which leads to a different Shear-Alfv{\'e}n gap structure with increased continuum damping, either via a steeper $q$-profile or a very flat background density profile.
The resulting more localized modes would prevent radial resonance overlap and lead to radially well separated redistribution.
Comparable, but slightly weaker domino-like behaviour was found in the self-consistent nonlinear case with full damping and EP density scaled to 200\%. This scaling can be motivated by the strong NBI drive which is expected to approximately double the drive w.r.t.\ to a pure $\alpha$ particle population. However, since the NBI population differs significantly from the $\alpha$ particle population, a multi-species analysis is needed to give a more realistic statement.\\
This work intended to demonstrate the probability, that a domino-like effect in EP transport can be relevant for \iter. However, due to the limitations of the model, no predictions can be made yet. The most important effects that should be investigated further are the NBI EP population and the sensitivity of different -- including radius-depended -- velocity distribution functions. At this point of the study, the domino-like depletion in the nonlinear simulation shows, that for certain ``worst-case'' scenarios, it might not be sufficient, to investigate EP redistribution on a quasi-linear basis only, without taking into account non-local and multi mode effects. Even if (on the transport time scale), the background density and current (thus, $q$) profiles are leading to a stable situation, small changes (towards more box-like shape) in either background density or current can lead to``overshooting'' domino-like EP redistribution and TAE drive. The final situation might even oscillate around the marginal stable profile.

 \section{Outlook}
In the presented work, the EP population generated by neutral beam injection (NBI) is not taken into account yet. Due to the strong anisotropy of the NBI ions in velocity space, the phase space resonance regions for these particles are different. The NBI drive depends also on the mode number $n$ in a different way w.r.t.\ to the $\alpha$ particle drive. Further, the radial mode structures might slightly change if the NBI EP population is considered. Since a multi-species approach has already been implemented into \hagis\ and is in the phase of testing, a follow-up publication will present an investigation including the NBI population.
 
The presented work raises the question, if a large amount of EP losses could be expected for this (or other) \iter\ scenarios. Since the redistribution in the worst-case reaches out to $s\approx 0.8$, possible interaction with the 3D field ripple perturbation has to be taken into account. A collaboration with the \ascot\ code \cite{Hirvijoki14} simulating the edge-localized field ripple with given \ligka\ perturbations is already ongoing. It allows for a first, preliminary estimate, that the amount of EP losses would not be dangerously large, since the loss regions calculated by \ascot\ and the radial extension of high amplitude areas in the \hagis\ simulation barely overlap. However also other AEs  (here only TAEs) and global MHD modes such as islands, have to be included.\\
Beside the question of real losses, it is planned in the near future to investigate the physical mechanisms and relevant time scales of particle redistribution. Especially it will be studied whether EPs are moving from the center to the edge on the basis of resonant or diffusive processes in the different stages of multi mode-particle interaction. For that study, the newly implemented \hmt\ \cite{Briguglio14} will be used within \hagis\ in combination with prescribed amplitude evolution to lower the CPU costs of the simulation.\\

\footnotesize{
\section{Acknowledgment}
The authors would like to thank Fulvio Zonca and the team of the NLED Enabling Research project for fruitful discussions.\\
The simulations for this work were partly run on HYDRA and the local Linux cluster (both from Rechenzentrum Garching) and on HELIOS at IFERC, Japan. The computational resources as well as the support are gratefully acknowledged.\\
This work has been carried out within the framework of the EUROfusion Consortium and has received funding from the European Union Horizon 2020 research and innovation programme under grant agreement number 633053. The views and opinions expressed herein do not necessarily reflect those of the European Commission.\\
The support from the EUROfusion Researcher Fellowship programme under the task agreement WP14-FRF-IPP/Schneller is gratefully acknowledged.
}

\vspace{-0.3cm}

\bibliographystyle{bibstyle}
\bibliography{PPCF100725_article}

\end{document}